\documentclass[12pt]{article}

\usepackage{color} 

\usepackage{amsmath}
\usepackage{amssymb,epsfig}
\usepackage[T1]{fontenc}
\usepackage[numbers]{natbib}
\usepackage[justification=raggedright]{caption}
\usepackage{float}
\usepackage{booktabs}
\usepackage{tikz}
\usetikzlibrary{ matrix,      
	fit,         
	positioning, 
}

\usetikzlibrary{shapes,decorations,arrows,calc,arrows.meta,fit,positioning}
\tikzset{
	-Latex,auto,node distance =1 cm and 1 cm,semithick,
	state/.style ={ellipse, draw, minimum width = 0.7 cm},
	point/.style = {circle, draw, inner sep=0.04cm,fill,node contents={}},
	bidirected/.style={Latex-Latex,dashed},
	el/.style = {inner sep=2pt, align=left, sloped}
}

\usepackage{multirow}
\usepackage{subfig}
\usepackage{rotating}
\usepackage[toc,page]{appendix}
\usepackage{url}
\usepackage{tabulary}
\usepackage{ctable}
\usepackage{epsfig}
\usepackage{graphicx}
\usepackage{caption}
\usepackage{multirow}
\usepackage{amssymb}
\usepackage{subfig}
\usepackage{footnote}
\usepackage{booktabs,caption}
\usepackage{tabularx}
\usepackage{tablefootnote}
\usepackage{esvect}

\usepackage[flushleft]{threeparttable}

\usepackage{lineno}

\newcommand{\bea}{\begin{eqnarray}}
\newcommand{\eea}{\end{eqnarray}}
\newcommand{\bss}{\begin{singlespace}}
\newcommand{\ess}{\end{singlespace}}		
\newcommand{\nn}{\nonumber}

\newcommand{\bfL}{{\mathbf{L}}}

\newcommand{\bfgamma}{{\pmb{\gamma}}}

\newcommand{\expit}{{\text{expit}}}

\newcommand{\bfU}{{\mathbf{U}}}

\newtheorem{defi}{Definition}

\newtheorem{lem}[defi]{Lemma}

\usepackage{tabularx}

\usepackage{makecell}

\newcommand\independent{\protect\mathpalette{\protect\independenT}{\perp}}
\def\independenT#1#2{\mathrel{\rlap{$#1#2$}\mkern2mu{#1#2}}}

\newcommand{\qedwhite}{\hfill \ensuremath{\Box}}
\newtheorem{theorem}{Theorem}

  \usepackage{setspace}
\usepackage{authblk} 

\title{Adjusting for time-varying treatment switches  \\ in randomized clinical trials: \\ the danger of extrapolation and how to avoid it} 
\date{}
\author[1]{Hege Michiels}
\author[2]{An Vandebosch}
\author[1]{Stijn Vansteelandt}
\affil[1]{Department of Applied Mathematics, Computer Science and Statistics, Ghent University, Ghent, Belgium}
\affil[2]{Janssen R\&D, a division of Janssen Pharmaceutica NV, Beerse, Belgium}

\begin{document}
\maketitle
\noindent Correspondence: {\tt hege.michiels@ugent.be}
	
\section*{Abstract}

When choosing estimands and estimators in randomized clinical trials, caution is warranted- as intercurrent events, such as - due to  patients who switch treatment after disease progression, are often extreme. Statistical analyses may then easily lure one into making large implicit extrapolations, which often go unnoticed. We will illustrate this problem of implicit extrapolations using a real oncology case study, with a right-censored time-to-event endpoint, in which patients  can cross over from the control to the experimental treatment after disease progression, for ethical reasons. We resolve this by developing an estimator for the survival risk ratio contrasting the survival probabilities at each time $t$ if all patients would take experimental treatment with the survival probabilities at those times $t$ if all patients would take control treatment up to time $t$, using randomization as an instrumental variable to avoid reliance on no unmeasured confounders assumptions.  This doubly robust estimator can handle time-varying treatment switches and right-censored survival times. Insight into the rationale behind the estimator is provided and the approach is demonstrated by re-analyzing the oncology trial.

 \noindent \emph{Keywords}:  Causal inference, Estimand, Hypothetical estimand, Instrumental variable, Treatment switching
 
\section{Introduction}


  Since the publication of the addendum of the ICH E9 guideline \cite{ICH2019}, the choice of estimands for treatment efficacy in the presence of intercurrent events has received more attention. The addendum is definitely a step in the right direction, as it emphasizes the importance of estimating an effect that answers the clinical question of interest. However, so far, relatively little attention has been paid to  criteria on the basis of which to choose an estimand, which estimators can be used for it,  and the assumptions on which they rely. Caution is nonetheless warranted when choosing estimands and estimators, as intercurrent events, such as - patients who switch treatment after disease progression, are often extreme. Therefore, statistical analyses may easily lure one into making large implicit extrapolations. In this paper, we will illustrate this problem using an oncology case study performed by Janssen Pharmaceuticals and propose solutions.

		In oncology trials, it is common that patients randomized to the control group can cross over to the experimental group during the trial \cite{halabi2019textbook, ishak2014methods,   jonsson2014analyzing,latimer2014adjusting}, typically after disease progression, defining a second line treatment \cite{graffeo2019ipcwswitch}. It can also occur due to changes in the treatment guidelines during the trial \cite{yingstructural}. 
		 For example, in  the HELIOS trial \cite{fraser2020final} (NCT01611090), performed by Janssen Pharmaceuticals,
		 578 patients with
		relapsed/ refractory chronic lymphocytic leukemia/small lymphocytic lymphoma without deletion
		17p  were randomized 1:1 to 420mg daily ibrutinib or placebo plus 6 cycles 
		of bendamustine plus rituximab (BR), followed by ibrutinib or placebo alone. Randomization was stratified  by purine analog refractory status (failure to respond or relapse in $\leq$ 12 months) and prior lines of therapy (1 line versus $>$1 line).
		Upon disease progression, patients from the placebo plus BR arm were allowed to cross over to ibrutinib, as an ethical requirement to offer experimental treatment to all. The median follow-up time was 63.7 months. At the final 5-year analysis, the intention-to-treat effect on the overall survival endpoint showed a significant effect from ibrutinib plus BR versus placebo plus BR (HR 0.611 95\% CI [0.455; 0.822]), despite crossover in 63.3\% of the control patients.   These kinds of scenarios with large proportions of treatment switching are not uncommon (e.g. \cite{munir2019final, feyerabend2019adjusting, fraser2020final, demetri2012complete, sternberg2013randomised}).

		While  patients may benefit from changing treatment,  it complicates the interpretation of the intention-to-treat (ITT) or treatment policy analysis \cite{halabi2019textbook}, which contrasts the treatment groups as randomized.  In particular, the magnitude of the ITT effect might have limited relevance as it  compares immediate experimental treatment with delayed experimental treatment \cite{halabi2019textbook}. However, adjusting for crossover is challenging as oncology trials often have complex designs, with different treatment changes at different points that may occur over time \cite{halabi2019textbook}. Moreover, a diversity of patient journeys is expected in oncological trials as patients can discontinue  treatment as a result of an adverse reaction, start a new anticancer regimen before observing the endpoint of interest or die while on treatment \cite{degtyarev2019estimands}.  In addition, whether it is useful to adjust for treatment switches depends on the clinical questions of interest, which are not always straightforward to define. 
		For example, if interest lies in the comparison of the effect of the two treatment policies implemented in the trial, then an intention-to-treat (ITT) or treatment policy estimand is ideal. However, a decision problem might also require an evaluation of what would have happened in the study if there were no treatment changes, corresponding to a hypothetical estimand. In general, it is important to define treatment switches clearly and specify which treatment changes one wishes to adjust for  \cite{halabi2019textbook}. 
  
		Different estimands and estimators have been used to correct for crossover as the treatment policy estimand is typically diluted in these settings \cite{ latimer2015adjusting, latimer2014adjusting,morden2011assessing, watkins2013adjusting}.
		Traditional methods include 
		removing switchers from the analysis, i.e. estimating a per-protocol effect, censoring patients at the time of crossover, or modeling the treatment as a time-varying covariate   \cite{graffeo2019ipcwswitch, yingstructural}. 
		 These methods are often used but are prone to selection bias as patients who cross over and those who do not are generally not comparable in terms of the survival times  one would observe if patients stayed on the assigned treatment \cite{yingstructural}. These methods should not be considered as providing conclusive evidence as they give up on the balance offered by randomization \cite{latimer2017adjusting}. 
		

	  In the HELIOS trial,  crossover does not reflect clinical practice as the experimental treatment was not yet approved at the time of conducting the trial. Therefore, 
		according to Manitz et al. \cite{manitz2022estimands}, the analysis should ideally target a hypothetical estimand \cite{ICH2019}, quantifying  the treatment effect if none of the patients had switched treatment.  In this paper,  we will focus on the estimation of this hypothetical estimand. 
	
		Several estimators have been developed to  estimate hypothetical estimands. 
		For example, g-estimation methods under rank preserving structural failure time models (RPSFTMs) can be used to estimate the treatment effect under perfect compliance with the assigned treatment. These methods rely on randomization as an instrumental variable \cite{halabi2019textbook, walker2004parametric}, thus assuming that    (i) it is associated with  the treatments actually taken, (ii) it does not share a common cause with the survival times, and (iii) it only affects the survival times through treatments actually taken. These conditions are expected to hold in a double-blinded randomized trial and  will be formally introduced later on in the paper. Instrumental variable methods succeed to estimate the average effect of the treatments actually taken on the outcome, regardless of whether we measured the (time-varying) confounders between treatment and outcome by explicitly exploiting the randomization of study arms \cite{hernan2006instruments}.  G-estimation methods for RPSFTMs rely on those instrumental variable assumptions and additionally on a common treatment effect assumption that 
		 the treatment effect received in the experimental group does not differ from the treatment effect received by control patients who cross over to the experimental treatment. 
		 A major limitation of these methods is that the handling of administrative censoring requires artificially censoring the observed event times for some patients. A further key concern is that it is vulnerable to extrapolation due to the parametric nature of the model.  In particular,  it is assumed that the experimental treatment extends the underlying survival by a constant amount, regardless of when the treatment is given \cite{watkins2013adjusting}.


	   The inverse probability of censoring weights (IPCW) method \cite{robins2000correcting} instead censors patients at the time of crossover and reweights  the remaining records to remove selection bias. It does not suffer from  the aforementioned disadvantages. 
	   However, it relies on the assumption that, at each time, there are no systematic differences between those who switch and those who do not, conditional on  measured variables, i.e.  there should be no unmeasured confounding between crossover and the outcome \cite{halabi2019textbook}.  Often only baseline covariates are included in the IPCW analysis, violating this assumption if switching is based on time-varying confounders like disease progression, as is nearly always the case.  In IPCW analyses, it is therefore essential to include   disease progression  as a time-varying predictor in the inverse probability weights. Even when this is done the IPCW estimator shows erratic behavior  in case of perfect prediction for crossover. In that case it can suffer from so called positivity violations \cite{halabi2019textbook} as there are no similar patients, in terms of the measured confounders, who cross over and who do not.  In that case certain subgroups, defined by the measured confounders, would always cross over to experimental treatment if assigned to the control arm.  Such positivity violations are common with treatment crossover because study protocols may indicate when, based on measured confounders,  patients should cross over,   and are problematic as they can lead to significant bias, large variance and invalid inference \cite{petersen2012diagnosing,  rudolph2022effects}.  Violations of this assumption in inverse weighting analyses will  typically be reflected in  large weights and are therefore generally easy to diagnose. However, this  can go unnoticed when weights are truncated. 
	   In addition, imputation strategies hide this problem by making implicit extrapolations \cite{vansteelandt2011invited}. 
	   
	   In the final analysis of the HELIOS trial,  only an intention-to-treat estimand was considered but corrections for crossover  have been made   in  interim \cite{chanan2016ibrutinib} and 3-year analyses \cite{fraser2019updated}.  In particular, inverse probability of censoring weights and rank preserving structural failure time models were used in an attempt to target the treatment effect one would have observed if patients would have stayed on the assigned treatment. 
	   As in many such analyses, it is unclear what covariates have been used to correct for crossover in the IPCW analysis and which model estimation method was used  in the RPSFTM analysis. 
	   This makes  it  impossible to judge whether the assumptions made are plausible.  In addition, extrapolations are a major complication of these analyses as patients switch after disease progression, implying that there are no comparable patients, in terms of measured confounders, who cross over and who do not. 

	 To alleviate concerns about extrapolation, it is often wise to target less ambitious estimands. 
For instance, Rudolph et al. \cite{rudolph2022effects} discussed positivity violations in  health policy evaluation settings and proposed to  redefine the estimand to correspond to a shift intervention or a modified treatment policy \cite{diaz2021nonparametric, haneuse2013estimation, munoz2012population}.  Michiels et al. \cite{michiels2021novel} also proposed an alternative estimand to lessen the extrapolations that are typically needed to infer the treatment effect that would have been observed had all patients stayed on the assigned treatment.
Alternatively,  a solution may be obtained in terms of estimators rather than estimands. In particular, in this paper, we develop an instrumental variable estimator for the hypothetical risk ratio, comparing always taking experimental treatment  to always taking control treatment, that does not rely on no unmeasured confounders assumptions. 
Our method does not suffer from the same concerns with censoring as an RPSFTM analysis, as a result of modeling survival probabilities instead of event times. It
 builds on the work by Ying and Tchetgen Tchetgen \cite{yingstructural} but has the advantage that it preserves the $p$-value from the treatment policy analysis. We consider this as an advantage, as under the null hypothesis of no treatment effect, crossover has no impact on the survival times and therefore the $p$-value obtained by a treatment policy analysis is valid. This may sound disadvantageous but is reflective of the fact that `honest' IV analyses extract information from the data and not from structural assumptions. In contrast,  IPCW analyses can achieve
 more power by relying on a no unmeasured confounders assumption and a  model assumption to predict crossover.

In this paper, we consider the setting where patients can cross over from the control to the experimental arm. We aim to provide insight into the rationale behind the model  proposed by Ying and Tchetgen Tchetgen 
using simple examples and figures, so that it can be applied in case studies.  Moreover, we improve the estimator proposed by Ying and Tchetgen Tchetgen by developing a doubly robust estimator, relying on a model for the randomized arm and a model for the hazards of death, that is unbiased even if one of these models is misspecified. By additionally using a model for the hazard, efficiency gains in the treatment effect estimator are expected. 
 In addition, software in \texttt{R} to apply our method is provided. 
The approach is illustrated by re-analyzing the HELIOS trial.

\section{Estimands}

\begin{figure}[h]
	\captionsetup[subfigure]{}
	\centering
	\subfloat[Hazard ratio: 0.30 (95\% CI: ${[0.18; 0.29]}$)]{\includegraphics[width=0.5\textwidth]{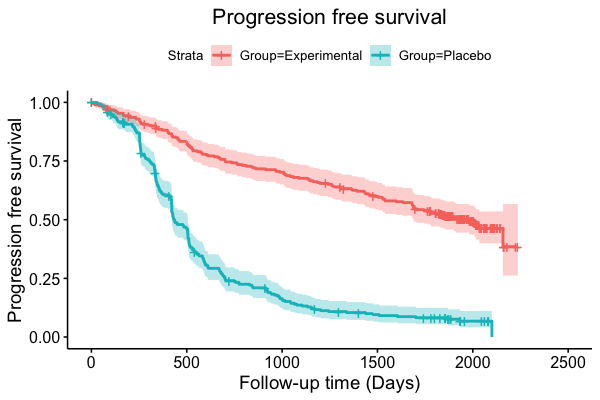}} 
	\subfloat[Hazard ratio: 0.61 (95\% CI: ${[0.46; 0.82]}$)]{\includegraphics[width=0.5\textwidth]{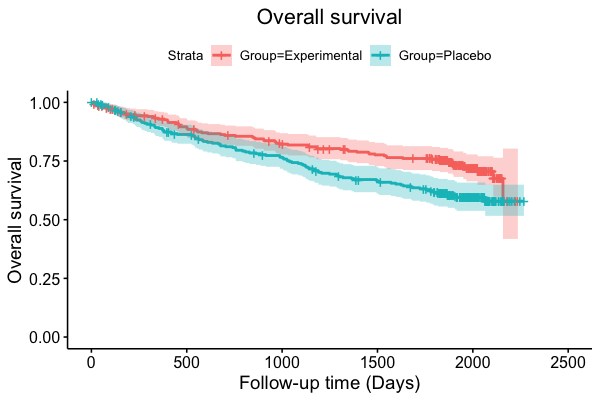}}  
	\caption{Kaplan-Meier curves of the progression-free and overall survival in the HELIOS trial, with 95\% confidence bands.} 
\label{Fig: HELIOS PFS OS}
\end{figure}

\begin{figure}[h]
	\centering
	\includegraphics[width=0.5\textwidth]{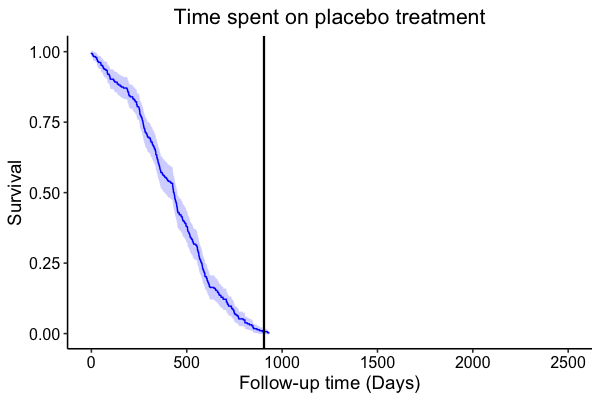} 
	\caption{Kaplan-Meier curve indicating the time patients in the control arm of the HELIOS trial spent on placebo treatment, with death as a competing event. 63\% of the control patients cross over to the experimental treatment. } 
	\label{Fig: HELIOS pb tmt}
\end{figure}

 In the HELIOS trial, the overall survival hazard ratio is diluted compared to the progression free survival hazard ratio (figure \ref{Fig: HELIOS PFS OS}), as many control patients start experimental treatment after disease progression, impacting the overall survival times. 
According to Manitz et al. \cite{manitz2022estimands}, in the HELIOS trial, it is recommended to  consider the hypothetical scenario in which crossover would not have existed, because crossover does not represent clinical practice given that the experimental treatment was not yet approved at the time of conducting the trial. 
However, estimating  the 5-year mortality risk if all control patients would stay on their assigned treatment requires strong extrapolations. This is because, after approximately 1000 days of follow-up, all control patients who were still in the trial stopped taking control treatment, crossed over to the experimental treatment, or started subsequent therapy (figure \ref{Fig: HELIOS pb tmt}).
Therefore,  inferring the effect if all patients would stay on the assigned treatment for 5 years is too ambitious, though not unusual to consider. 
In the remainder of the paper, we will therefore focus on estimating the hypothetical effect of experimental treatment up to the day of unblinding the trial had none of the patients crossed over from control to experimental, since we do have information about taking control treatment until that day.  This day is indicated with a black vertical line in figure 	\ref{Fig: HELIOS pb tmt}. In the remainder of the paper, for ease of explanation, we refer to this day as the end of the trial. 


\subsection{Hypothetical risk ratio}

In this paper, we develop an estimator that targets a hypothetical estimand considering the setting that all patients would stay on the assigned treatment. In particular, we consider the  risk ratio contrasting the survival probabilities if all patients would take experimental treatment 
with the survival probabilities if all patients would take control treatment.
This estimand can be evaluated at different time points in the trial, as it is of interest to observe how the effect evolves over time. In the next section, this estimand, together with an instrumental variable estimator, is formally introduced. 

\section{Instrumental variable estimator}
\label{Section: IV estimator}

Ying and Tchetgen Tchetgen \cite{yingstructural} proposed an estimator for a causal treatment effect to account for selective treatment switching using a structural nested cumulative survival time model (SNCSTM)  for censored time-to-event outcomes \cite{seaman2020adjusting}.  As in the Aalen additive hazards model \cite{aalen1989linear}, it is assumed that the treatment has an additive effect on the hazard, unlike Cox proportional hazard models \cite{cox1972regression} that assume a multiplicative effect of the treatment on the hazard. However, while the Aalen additive hazards model can be used to estimate the treatment policy effect of being assigned to the experimental arm compared to being assigned to the control arm, the SNCSTM can be used to target the hypothetical effect if all patients would stay on the assigned treatment for the entire duration of the trial. 
This SNCSTM is fitted using randomization as an instrumental variable, avoiding unconfoundedness or positivity assumptions. However, the estimator for this SNCSTM proposed by Ying and Tchetgen Tchetgen has the disadvantage of not preserving the $p$-value obtained in a treatment policy analysis. Therefore, in this paper, we improve the estimator proposed by Ying and Tchetgen Tchetgen \cite{yingstructural} by using a doubly robust estimator, relying on a model for the randomized arm and a model for the hazards of death, that is unbiased even if one of these models is misspecified. Moreover, by adjusting for baseline covariates we require a weaker assumption for administrative censoring or censoring due to drop-out. 
 Readers who are not interested in the technical details of the estimand and estimator can skip this section and continue reading from section \ref{Section: Data analysis} onwards. 
 
\subsection{Notation}

\begin{figure}[H]
	\centering
	\subfloat[]{\includegraphics[width=0.75\textwidth]{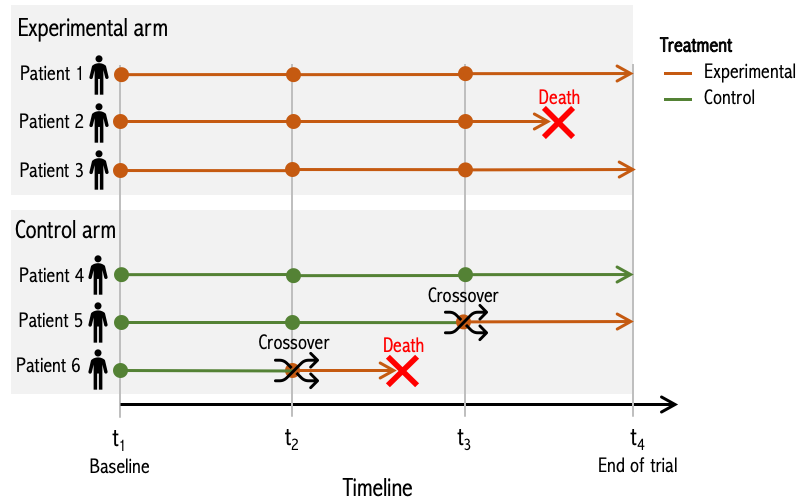}}  \\
	\subfloat[]{ \begin{tabular}{ccccccc}\hline
					Patient & Arm & $Z$ &\multicolumn{4}{c}{Treatments} \\ 
				& & & $D(t_1)$ & $D(t_2)$ & $D(t_3)$ & $D(t_4)$  \\ \hline
				1 & Experimental & 0 & 0&0&0&0 \\
				2&Experimental & 0 & 0&0&0&/\\
				3&Experimental & 0 & 0&0&0&0\\ 
				4& Control & 1 & 1&1&1&1\\
				5& Control & 1 & 1&1&0&0\\
				6& Control & 1 & 1&0&/&/\\ \hline
	\end{tabular}}
	\caption{(a) Illustration of different possible patient pathways in the setting where patients can cross over from the control to the experimental arm, (b) with corresponding notation.} 
	\label{Fig: Patient pathways}
\end{figure}

Let $\tilde{T}$ denote the survival time, $C$ the potential censoring time and $T = \min(\tilde{T},C)$ the censored survival time. Let $\delta = I(\tilde{T}\leq C)$ be the event indicator, $N(t) = I(T\leq t, \delta=1)$ the counting process and $Y(t) = I(T\leq t)$ the at-risk process.
 In addition, we denote the randomized arm by $Z$, with 1 indicating the control arm and 0 the experimental arm,  and the vector of baseline covariates by $\bfL$. Let $D(t)$ indicate the treatment taken at time $t$, with $D(t) =0$ if the patient takes experimental treatment at time $t$ and 1 otherwise. We assume that the treatments are taken at discrete times $t \in \{t_1, \dots, t_M\}$. The history of treatments is denoted by a bar, i.e. $\overline{D}(t_m) = (D(t_1), D(t_2), \dots, D(t_m))$ for $t_m \in \{t_1, \dots, t_M\}$. The treatment $ D(t_m)$ is not observed if $t_m > T$. It is assumed that the first treatment is taken at baseline, i.e. $t_1$ corresponds to the baseline visit. Consequently, for all patients in the trial, we have $T\geq t_1$.  The causal ordering of the variables is $\{ T \wedge t_1, D(t_1), T \wedge t_2,  D(t_2) ,\dots, T \wedge t_M,D(t_{M})\}$, where $x \wedge y$ indicates the minimum of $x$ and $y$.  The observed data consist of independent and identically distributed observations $\{(Z_i, \bfL_i , \overline{D}_i(T_i), T_i, \delta_i), i = 1, \dots, n\}$.  Different patient pathways and notation are illustrated in figure \ref{Fig: Patient pathways}.

In the SNCSTM \cite{yingstructural},  counterfactual survival times for time-varying treatments \cite{robins1986new, robins1987graphical} are modeled. Let $\tilde{T}(1_m, 0)$ denote the potential time to event had the patient, possibly contrary to fact,  taken control treatment up to time $t_m$ and experimental treatment thereafter. 
It holds that $T = {T}(1_m, 0)$ for patients who take control treatment up to time $t_m$ and experimental treatment thereafter. In addition,  intervening on treatment can only impact the survival after the time of that treatment, i.e. the event $\tilde{T}(1_{m-1}, 0)\geq t_m$ only occurs if the event $\tilde{T}(1_l, 0)\geq t_m$ occurs for all $l\geq m$ \cite{yingstructural}. Consequently, $\tilde{T} \geq t$ and $\tilde{T}(1_m, 0)\geq t$ are the   same events for $t \in [t_m, t_{m+1}[$.

\subsection{Model}

\begin{figure}[H]
	\centering
	\includegraphics[width=0.8\textwidth]{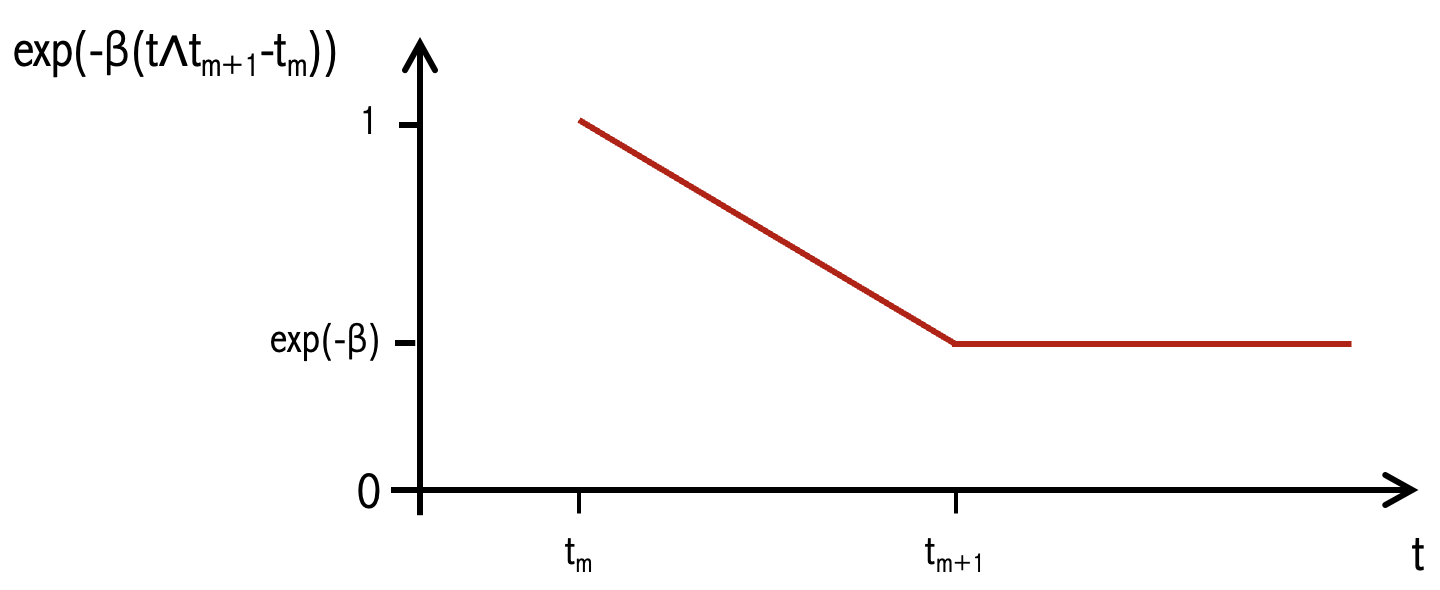}
	\caption{Illustration of SNCSTM (\ref{eq: SNCSTM general}), for $\beta$>0. }
	\label{fig: model illustration}
\end{figure}

In this section, we discuss the SNCSTM \cite{martinussen2011estimation}, contrasting the ratio of survival probabilities at time $t$ upon starting experimental treatment at time  $t_m \leq t$ versus at time $t_{m+1}$, for patients still alive at time $t_m$,  in the control arm, taking control treatment till time $t_{m}$ and with baseline covariates $\bfL$:
\begin{equation}
	\frac{P[\tilde{T}(1_m, 0)>t  | \overline{D}(t_m)= \overline{1}, Z=1, \bfL, \tilde{T}\geq t_m]}{ P[\tilde{T}(1_{m-1}, 0)>t  | \overline{D}(t_m)= \overline{1}, Z=1, \bfL, \tilde{T}\geq t_m]} 
	= \exp\left\{-  \beta  (t\wedge t_{m+1} - t_m) \right\},
	\label{eq: SNCSTM general} 
\end{equation}
with unknown scalar parameter $\beta$. In figure \ref{fig: model illustration}, this model is illustrated, showing that it is assumed that the effect of taking experimental treatment  decreases till the next visit and stays constant afterwards.  This is realistic as a new dose of treatment is taken at the next visit. 

By taking logs of each side of equation (\ref{eq: SNCSTM general}) and differentiating with respect to $t$, it can be shown that  this model can be written equivalently as
\begin{equation*}
	h_{\tilde{T}(1_m, 0)} (t | \bfL) -h_{\tilde{T}(1_{m-1}, 0)} (t | \bfL) = \beta I(t \leq t_{m+1} ),
\end{equation*}
where $	h_{\tilde{T}(1_m, 0)} (t | \bfL)$  and $h_{\tilde{T}(1_{m-1}, 0)} (t | \bfL)$  are the conditional hazards of $\tilde{T}(1_m, 0)$ and $\tilde{T}(1_{m-1}, 0)$  at time $t$ for control patients who were still  taking control treatment at visit $t_{m-1}$ and with baseline covariates $\bfL$. Note that, after time $t_{m+1}$, the hazard difference becomes 0, as it is assumed that the effect of taking experimental treatment at time $t_m$ stops at the next visit.  Thus, the effect of crossover at every time boils down to a decrease in the hazard of death with $\beta$ units. Therefore,   the experimental treatment is assumed to have an additive effect on the hazard and $\beta$ describes a hazard difference.  A positive value implies that the experimental  treatment increases patients' survival probability. This model is referred to as the  `constant hazards difference model' \cite{yingstructural}. In Appendix \ref{Section: Appendix SNCSTM}, a more general model is described, allowing the effect of experimental treatment on the hazard to change over time. 

\begin{figure}[H]
	\captionsetup[subfigure]{}
	\centering
	\subfloat[ ]{\includegraphics[width=0.5\textwidth]{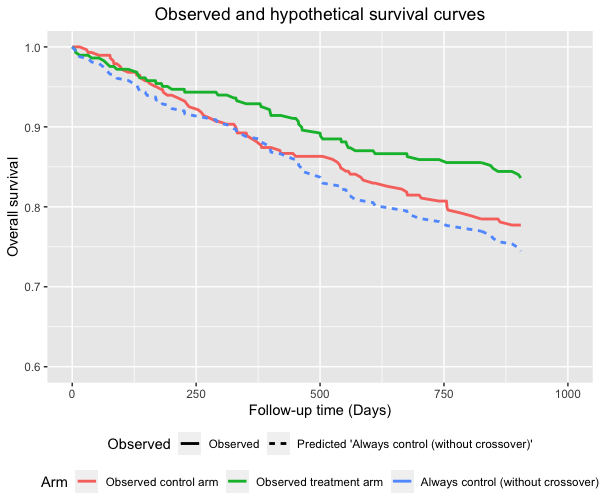}}
	\subfloat[]{\includegraphics[width=0.5\textwidth]{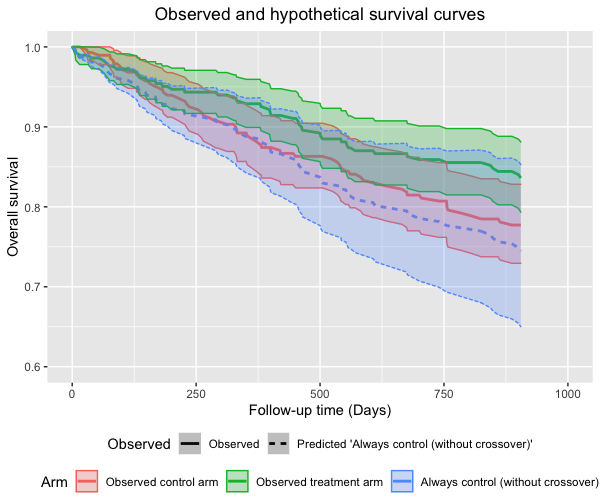}}
	\caption{(a) Illustration of observed experimental and control  Kaplan-Meier survival curves in the HELIOS trial and predicted control survival curves if nobody would crossover (b) with 95\% confidence interval. This curve is based on the estimated $\beta$ using the one-step estimator (see section \ref{Section: Data analysis}). }
	\label{fig: survival curves beta}
\end{figure}

Parameter $\beta$ can be used 
to target the hypothetical risk ratio comparing always treated versus never treated regimen up to time $t$  \cite{yingstructural} (see section \ref{Section: Hypothetical risk ratio}): 
\begin{equation}
	\frac{P[\tilde{T}(0)>t | \bfL]}{P[\tilde{T}(1)>t  | \bfL ]}  = \exp\left\{ \beta t\right\}. 
	\label{eq: hypothetical survival risk ratio1}
\end{equation}
Note that the model assumes this ratio not to depend on the baseline covariates $\bfL$.  As patients can only cross over from the control to the experimental arm and not vice versa, the counterfactual survival probability $P(\tilde{T}(0)>t| \bfL )$ equals the observed survival probability in the treatment arm $P(\tilde{T}>t| Z=0, \bfL)$, which can be estimated using a Kaplan-Meier estimator. In that case,  equation (\ref{eq: hypothetical survival risk ratio1}) can be used to predict the survival probabilities  $P(\tilde{T}(1)>t| \bfL )$ if patients would take control treatment and not cross over, for a given value of $\beta$ (see section \ref{Section: Hypothetical risk ratio}). This is illustrated in figure \ref{fig: survival curves beta}. 
An advantage of this   risk ratio is that it can be interpreted as a causal contrast, unlike hazard ratios  that have a built-in selection bias \cite{hernan2010hazards}.  The role of the baseline covariates $\bfL$ will be discussed in section \ref{Section: Hypothetical risk ratio}.

\subsection{Assumptions}

The randomization indicator $Z$ is used as an instrumental variable, meaning that it is assumed to  satisfy the following conditions \cite{yingstructural}: 
\begin{enumerate}
	\item \textbf{IV relevance } 
	
	The instrument is associated with the treatment actually taken at time $t_m$ for patients still at risk at time $t_m$, given history:  $Z  \not\!\perp\!\!\!\!\perp D(t_m)  |  \tilde{T} \geq t_m, \overline{D}(t_{m-1}), \bfL$, for all $t \in \{t_1, \dots, t_M\}$. 
	
	This assumption is stronger than strictly needed. It suffices to satisfy
	 an assumption expressed in terms of the expectation of  the derivative of the score function (see section \ref{Section: One-step estimator}). However, this assumption is less insightful. 
	
	\item \textbf{IV independence}
	
	The instrument is independent of the potential outcome under treatment, conditional on the baseline covariates:   $Z \independent \tilde{T}(0) | \bfL$.
		\item \textbf{Exclusion restriction}
	
The instrument has no direct causal effect on the outcome other than through the exposure: $\tilde{T}(\overline d, z) = \tilde{T}(\overline d, z')$, for all values of $\overline d$, $z$, and $z'$, where $\tilde{T}(\overline d, z)$ indicates the potential survival time had one set the treatments $\overline{D} = (D(t_1), D(t_2), \dots, D(t_M))$ to $\overline{d}$  and the treatment arm $Z$ to $z$. 
	
\end{enumerate}
In a double-blinded randomized trial, these assumptions are expected to hold. In particular, assumption 1 is satisfied if patients still at risk are more likely to receive treatment at time $t_m$ if they were assigned to treatment, even after conditioning on treatment and covariate history. This is especially expected to hold in the HELIOS trial as patients can only cross over from the control to the experimental arm. The second assumption is   expected to hold because of  randomization and the third because of  double-blinding. However, this could be violated if patients become aware of the treatment they are taking, e.g. because of adverse events.  This might be the case in the HELIOS trial as 69\% of the patients in the experimental arm experienced a serious treatment-emergent adverse event \cite{fraser2020final}.

We also make the following assumption about administrative censoring or censoring due to drop-out: 
\begin{itemize}
	\item[4.] \textbf{Conditional independent censoring}
	
	Censoring is independent of survival times, randomization and treatments actually taken, given baseline covariates:  $C\independent (\tilde{T},Z, \overline{D}(t)) |\bfL $. 
\end{itemize}
 This assumption can be relaxed by using inverse probability of censoring weighting \cite{robins2000correcting} but we will not discuss this further here.

\subsection{Identification}

\subsubsection{Estimator by Ying and Tchetgen Tchetgen}

Ying and Tchetgen Tchetgen \cite{yingstructural} proposed a two-step estimator for the parameter of interest $\beta$. In the first step, a model for the expected assigned treatment given baseline covariates, i.e. $E(Z| \bfL)$, is fitted. Note that this probability is known by design in a randomized trial.  
In the second step, an estimator of $\beta$ is obtained using an explicit recursive form. 
Ying \cite{ying2022ivsacim} developed an \texttt{R} package called `\texttt{ivsacim}' that implements this two-step estimator. Moreover, an estimator for a more general model, allowing the effect of experimental treatment on the hazard to change over time, together with a goodness-of-fit test for the constant hazards model, is provided. 
However, at the moment of writing, the package does not allow to include baseline covariates in the propensity score model for $E(Z| \bfL)$.  
By doing so, it is assumed that censoring due to drop-out or administrative censoring is marginally independent of survival times, randomization and treatments actually taken, which is possibly a much stronger assumption than our "conditional independent censoring" assumption. 
 
 Ying and Tchetgen Tchetgen  \cite{yingstructural} proved that their proposed estimator is uniformly consistent and asymptotically normal. However, the $p$-value of $Z$ in the Aalen additive hazards model \cite{aalen1989linear}, targeting a treatment policy estimand, is not preserved.  This can be problematic as different conclusions may then be drawn based on a treatment policy estimand than based on the hypothetical estimand.
   Therefore, in  section \ref{Section: One-step estimator}, we introduce a new estimator that does preserve the $p$-value and has the advantage of being doubly robust. By adding a model for the hazard, we expect to obtain a more efficient treatment effect estimator. Moreover, model selection can be performed for this model \cite{chernozhukov2018double}.
 First, we briefly describe the Aalen least squares estimator \cite{aalen1989linear} for the treatment policy estimand.

\subsubsection{Aalen least squares estimator for the treatment policy estimand} 
\label{Section: Aalen least squares estimator}

Consider a semiparametric additive hazards model \cite{mckeague1994partly} with time-varying intercept $ \delta(t)$ and constant treatment effect $\beta_A$: 
\begin{equation}
	\lambda(t | Z) =   \delta(t) + \beta_A Z .  
	\label{eq: aalen model}
\end{equation}
In this model, $\beta_A$ represents a treatment policy effect. In particular, it is 
the additive effect on the hazard of being assigned to the experimental arm compared  to being assigned to the control arm. 
An estimator for  $\beta_A$ can be obtained by solving estimating equations 
\begin{eqnarray}
	\frac{1}{n} \sum_{i=1}^{n} 	\bfU(Z_i,T_i;\beta_A) = 0, 
	\label{eq: estimating equations Aalen}
\end{eqnarray}
with 
\begin{eqnarray}
		\bfU(Z,T;\beta_A) \nn  =   \int_{0}^{\tau} \left\{Z-	E(Z | T\geq t ) \right\} \left\{ dN(t) -  \beta_AY(t)dt  - Y(t)\lambda(t)dt \right\},	\label{eq: estimating equation Aalen}
\end{eqnarray}
with $\tau$ the time point indicating the end of the study. A $p$-value for the test $\beta_A = 0$ can be obtained by performing a one-sample t-test on the score functions $U(Z,T;\beta_A = 0) $ evaluated in $\beta_A = 0$ and the variance of $\hat{\beta}_A$  using the sandwich estimator. 
In addition, a 95\% confidence interval for $\beta_A$ can be found by searching for the values that lead to  a $p$-value of 5\% in the t-test on the score functions.  Aalen additive hazards models, e.g. model (\ref{eq: aalen model}), can be fitted in  \texttt{R} using the \texttt{aalen} function from the \texttt{timereg} package. However, $p$-values and variance estimates provided by this function correspond to Wald tests rather than score tests. We will use score tests as the estimator for the hypothetical effect, discussed in the next section, which preserves the $p$-value provided by the treatment policy analysis, by imitating its score equations.

\subsubsection{One-step estimator}
\label{Section: One-step estimator}

We developed a one-step estimator \cite{le1956asymptotic} for the hypothetical effect $\beta$, where the initial estimator is obtained using the method by Ying and Tchetgen Tchetgen \cite{yingstructural}. This estimator is updated using a single Newton step on the score equation.  In particular, in Appendix \ref{Section: Appendix One-step estimator for constant treatment effect } we show that, for $t\geq 0$,
\begin{equation}
	\small
  E \left[ \left\{Z-E(Z | \tilde{T}(0) \geq t, \bfL   ) \right\} \exp\left\{\beta \int_{0}^{t}D(s)ds \right\}   \left\{dN(t) -\beta Y(t)D(t) dt -Y(t)\lambda(t|  Z=0, \bfL)dt  \right\} \right] = 0, 
  \label{eq: estimating equation}
\end{equation}
with 
\begin{equation*}
	E(Z | \tilde{T}(0) \geq t, \bfL) = \frac{E\left[Z Y(t)\exp\left\{\beta \int_{0}^{t}D(s)ds \right\} |\bfL  \right]}{E\left[Y(t)\exp\left\{\beta \int_{0}^{t}D(s)ds \right\} |\bfL  \right]}. 
	\label{eq: Z term}
\end{equation*}
In equation (\ref{eq: estimating equation}), the treatment effect at every time $t$ is subtracted from the counting and the at-risk process, thereby mimicking the counterfactual survival time when  always taking experimental treatment, i.e. $\tilde{T}(0)$.  The parameter $\beta$ is then chosen such that  the IV assumption  $Z \independent \tilde{T}(0) | \bfL$ is satisfied. However, in equation (\ref{eq: estimating equation}), we calculate the expected value of $Z$ conditional on $\tilde{T}(0)$ and $\bfL$,  to imitate the estimating function of the Aalen additive hazards model for the treatment policy estimand (\ref{eq: estimating equations Aalen}). 
 
From (\ref{eq: estimating equation}) it follows that,  if $\tau$ is the time point indicating the end of the study,  $\beta$ can be estimated by solving estimating equations 
\begin{eqnarray}
	\frac{1}{n} \sum_{i=1}^{n} {U}(Z_i,T_i,\overline{D_i}(\tau),\bfL_i;\beta) = 0, 
	\label{eq: estimating equations method 1}
\end{eqnarray}
with 
\begin{eqnarray*}
&&{U}(Z,T,\overline{D}(\tau),\bfL;\beta) \\
&&=   \int_{0}^{\tau} \left\{Z-	\hat{E}(Z | \tilde{T}(0) \geq t, \bfL) \right\} \exp\left\{\beta \int_{0}^{t}D(s)ds \right\}  \left\{ dN(t) -  Y(t) (\beta D(t) +\lambda(t|  Z=0, \bfL))dt \right\},   
\end{eqnarray*}
where $	\hat{E}(Z | \tilde{T}(0) \geq t, \bfL)$ is an estimate of ${E}(Z | \tilde{T}(0) \geq t, \bfL)$ obtained by fitting a logistic regression model, with $Z$ as outcome,  baseline covariates $\bfL$ as predictors and weights $Y(t)\exp\left\{\beta \int_{0}^{t}D(s)ds \right\}$.  Parameter $\beta$ can be estimated by solving estimating equations (\ref{eq: estimating equations method 1}) directly, but we propose a one-step estimator  \cite{le1956asymptotic}, which is usually computationally more efficient.  By starting from a consistent estimator for $\beta$, we obtain an asymptotically equivalent estimator (see theorem 5.45 in \cite{van2000asymptotic}, which also holds for M-estimators).  This motivates the following approach:
\begin{enumerate}
	\item Compute an initial estimate $\hat{\beta}_0$ for $\beta$ using the method by Ying and Tchetgen Tchetgen \cite{yingstructural}. 
	\item Fit an Aalen additive hazards model or Cox proportional hazards model in the experimental arm for the observed event times, conditional on baseline covariates $\bfL$.
	\item Let $\tau_1, \dots, \tau_k$ denote the observed event times in the dataset. For every $\tau_j$ ($j \in \{1, \dots, k\}$): 
	\begin{enumerate}
		\item Use the model from step 2 to estimate hazards $\lambda(\tau_j| Z = 0, \bfL)$ for every patient.  
		\item Fit a  weighted logistic regression model with $Z$ as outcome,  baseline covariates $\bfL$ as predictors and weights $Y(\tau_j)\exp\left\{\hat{\beta}_0 \int_{0}^{\tau_j}D(s)ds \right\}$.  As patients can only cross over from the control to the experimental arm, these weights simplify to $Y(\tau_j)$ for patients in the experimental arm,  $Y(\tau_j) \exp\left\{\hat{\beta}_0 \tau_j \right\}$	for control patients who did not cross over before time $\tau_j$ and $Y(\tau_j)\exp\left\{\hat{\beta}_0 S_i \right\}$ for control patients who did cross over at time $S_i < \tau_j$. 
		\item Use the model from step 3 (b) to make predictions for every patient. Denote these predictions as $\hat{E}(Z | \tilde{T_i}(0) \geq \tau_j, \bfL_i)$ ($i \in \{1, \dots, n\}$).  
		\item Calculate 
		\begin{align*}
			{U}_{\tau_j}(Z_i,T_i,\overline{D}_i(\tau),\bfL_i;\beta_0) 
			&=  \left\{Z_i-	\hat{E}(Z | \tilde{T}_i(0) \geq \tau_j, \bfL_i) \right\} \exp\left\{\beta_0 \int_{0}^{\tau_j}D_i(s)ds \right\} \\
			& \times   \left\{ dN_i(\tau_j) -  Y(\tau_j) \left\{\hat\beta_0D(\tau_j)+\lambda(\tau_j|  Z=0, \bfL) \right\}(\tau_j -\tau_{j-1} )  \right\},  
		\end{align*}
		with $\tau_0$ = 0,  for every patient. 
	\end{enumerate}
	\item Calculate the score function 
	${U}(Z_i,T_i,\overline{D}_i(t_M),\bfL_i;\hat\beta_0) = \sum_{j=1}^{k} {U}_{\tau_j}(Z_i,T_i,\overline{D}_i(\tau),\bfL_i;\hat\beta_0)$ for every patient.
	\item Calculate the derivative of the score function w.r.t. $\beta$, i.e. $\dot{U}( T_i, Z_i, \bfL_i, \overline{D}_i(t_M);\hat{\beta}_0)$,  for every patient, evaluated in $\hat{\beta}_0$, as described in Appendix \ref{Section: appendix Derivative of score equation}.
	\item Update the initial estimate of $\beta$: 
	\begin{equation*}
		\hat{\beta} = \hat{\beta}_0 - \sum_{i=1}^n {U}( T_i, Z_i, \bfL_i, \overline{D}_i(t_M); \hat{\beta}_0)/  \sum_{i=1}^n \dot{U}(T_i, Z_i, \bfL_i, \overline{D}_i(t_M); \hat{\beta}_0). 
	\end{equation*}
\end{enumerate} 
We implicitly assume that the expectation of the derivative of the score function differs from 0 when evaluated at the truth $\beta^*$, i.e. $E(\dot{U}( T, Z, \bfL, \overline{D}(t_M));\beta^*) \neq 0$,  which typically holds because of the IV relevance assumption.  The variance of $\hat{\beta}$ can be estimated using the sandwich estimator \cite{huber1967behavior}: $$\text{var}(\hat{\beta}) \cong \frac{1}{n} \text{var}({U}( T_i, Z_i, \bfL_i, \overline{D}_i(t_M));\hat{\beta})/E(\dot{U}(T_i, Z_i, \bfL_i, \overline{D}_i(t_M));\hat{\beta})^2.$$ A $p$-value for the test $\beta = 0$ can be obtained  by performing a one-sample t-test on the score functions ${U}( T_i, Z_i, \bfL_i, \overline{D}_i(t_M);\beta=0)$ evaluated in $\beta = 0$. 
In addition, the boundaries of a  95\% confidence interval for $\beta$ can be found by searching for the values that lead to a $p$-value of 5\% in the t-test on the score functions. In  the online supplementary material, it is illustrated how this can be implemented in \texttt{R}. 

As proven in Appendix \ref{Section: Appendix Proof of double robustness}, the obtained estimator for $\beta$ is  doubly robust, meaning that it is unbiased even if 
the model for the randomized arm, i.e. $E(Z | \tilde{T}(0) \geq t, \bfL)$, or the model for the hazards, i.e. $\lambda(t|  Z=0, \bfL)$, but not both,
is misspecified. 
By adding a model for the hazard, we expect to obtain a more efficient treatment effect estimator. Moreover, as the estimator is doubly robust, model selection can be performed for the model for the hazard \cite{chernozhukov2018double}. 

This estimating function for $\beta$ equals the estimating equation of the treatment policy estimand in the Aalen additive hazards model under the null hypothesis $\beta = 0$. Therefore, we will obtain the same $p$-value when performing a score test in the treatment policy analysis.

\subsection{Hypothetical risk ratio}
\label{Section: Hypothetical risk ratio}


Consider the  additional assumption of no-current treatment value interaction, stating that the causal effect of taking control  at time $t_m$ versus experimental treatment,  among patients who were taking control at time $t_m$ is equal to that among patients who were treated at time $t_m$ (see Appendix \ref{Section: Appendix Hypothetical estimand}).  Under this assumption,  the effect $\exp\left\{ \beta t\right\}$ can be interpreted as the hypothetical risk ratio comparing always treated versus never treated regimes up to time $t$  \cite{yingstructural}: 
\begin{equation}
\frac{P[\tilde{T}(0)>t | \bfL]}{P[\tilde{T}(1)>t  | \bfL ]}  = \exp\left\{ \beta t\right\}.
	\label{eq: hypothetical survival risk ratio}
\end{equation}
The no-current treatment value interaction assumption could be violated if patients who cross over do not experience the same treatment effect as those who do not. As these are two, possibly very different, subgroups, this assumption can be violated. However,   it can be relaxed by including time-varying covariates in the SNCSTM, but this complicates the estimation procedure. 
Implementation of this  hypothetical estimand  (\ref{eq: hypothetical survival risk ratio}) is illustrated in  the online supplementary material. 

As patients can only cross over from the control arm to the experimental arm and not vice versa, the survival probability if all patients would take control treatment and not cross over can be estimated using risk ratio (\ref{eq: hypothetical survival risk ratio}), as shown in Appendix \ref{Section: Appendix counterfactual survival curve under control}: 
\begin{eqnarray*}
	\hat{P}(\tilde{T}(1)>t ) 
	&&= \exp\{-\hat{\beta}t\} \frac{1}{n} \sum_{i=1}^{n}  \hat{P}(\tilde{T}>t | Z_i = 0, \bfL_i ). 
\end{eqnarray*}
The survival probabilities $\hat{P}(\tilde{T}>t | Z_i = 0, \bfL_i )$ can be obtained by fitting a Cox proportional hazards model or an Aalen additive hazards model to the observed experimental arm. 

To make the conditional independent censoring assumption more plausible, it is beneficial to add more baseline covariates in $\bfL$. However, in model (\ref{eq: SNCSTM general}), it is assumed that the ratios of survival probabilities do not depend on $\bfL$. Therefore, adding more covariates makes model (\ref{eq: SNCSTM general}) more restrictive. 


\section{Data analysis}
\label{Section: Data analysis}

\subsection{Methods}

In the data analysis of the HELIOS trial, we consider the treatment policy and hypothetical estimand regarding the intercurrent event `treatment crossover'.  In addition, we apply three traditional methods, i.e. excluding patients who cross over, censoring patients at the time of crossover and including treatment as a time-varying covariate. The hypothetical estimand is targeted using the IPCW method, the IV method by Ying and Tchetgen Tchetgen and the one-step estimator proposed in this paper.  
For all effects, the relative risk (RR) at the end of the trial is considered as summary measure.  In particular, the treatment policy estimand contrasts the survival probabilities if all patients  would be assigned to the experimental arm versus the control arm, while the hypothetical estimand contrasts the survival probabilities if all patients  would take experimental treatment versus control treatment for the entire study duration. 
These estimands can be defined according to the guidelines of the ICH E9(R1) guideline (see Appendix \ref{Section: Appendix Data analysis guideline}).

All analyses are performed using the software \texttt{R} (version 4.0.4). For the treatment policy estimand and the traditional methods, these risk ratios are assessed using the \texttt{aalen} function from the \texttt{timereg} package \cite{scheike2022package}, fitting an Aalen additive hazards model \cite{aalen1989linear}.  
To target the hypothetical estimand, we apply the IPCW method  to the Aalen additive hazards model. In particular,  the  \texttt{ipcw} function from the \texttt{ipcwswitch} package \cite{graffeo2021package} is used to compute stabilized IPCweights with baseline and time-varying predictors, as described by Graffeo et al.  \cite{graffeo2019ipcwswitch}.  As baseline covariates, we include the two stratification factors, i.e. purine analog refractory status (failure to respond or relapse in $\leq$ 12 months) and prior lines of therapy (1 line versus $>$1 line), sex and age, and as time-varying confounders, a binary variable indicating whether or not, at every day,  the patient had already disease progression. This method is compared to the method where we additionally include the time since progression as time-varying predictor.  Finally, we also compute weights without time-varying predictors, e.g. only baseline covariates, as described by Willems et al. \cite{willems2018correcting}. The weights are then incorporated into an  Aalen additive hazards model with $Z$ as predictor. 

To target the hypothetical risk ratio, we also consider different IV methods. In particular, the method developed by Ying and Tchetgen Tchetgen \cite{yingstructural} is applied by using the \texttt{ivsacim} function from the \texttt{ivsacim} package \cite{ying2022package}. This method is compared to the one-step estimator discussed in section \ref{Section: One-step estimator}, including the two stratification factors, sex and age as baseline covariates. \texttt{R} code for these estimators can be found in the online supplementary material. To make a fair comparison of the considered methods,  all $p$-values for the tests $\beta = 0$ are obtained by performing a one-sample t-test on the score functions evaluated in $\beta = 0$ (see Appendix \ref{Section: Appendix Score equations Aalen additive hazards model}). In addition,  the variance of $\hat{\beta}$ is obtained using the Sandwich estimator and 95\% confidence intervals are obtained by searching for the $\beta$ values that lead to a $p$-value of 5\% in the t-test on the score functions. Finally,  the relative risks  are estimated using expression  (\ref{eq: hypothetical survival risk ratio}) and a 95\% confidence interval is obtained using the same transformation of the bounds of the confidence interval for $\beta$. 

\subsection{Results}

\begin{figure}[H]
	\centering
	\includegraphics[width=1\textwidth]{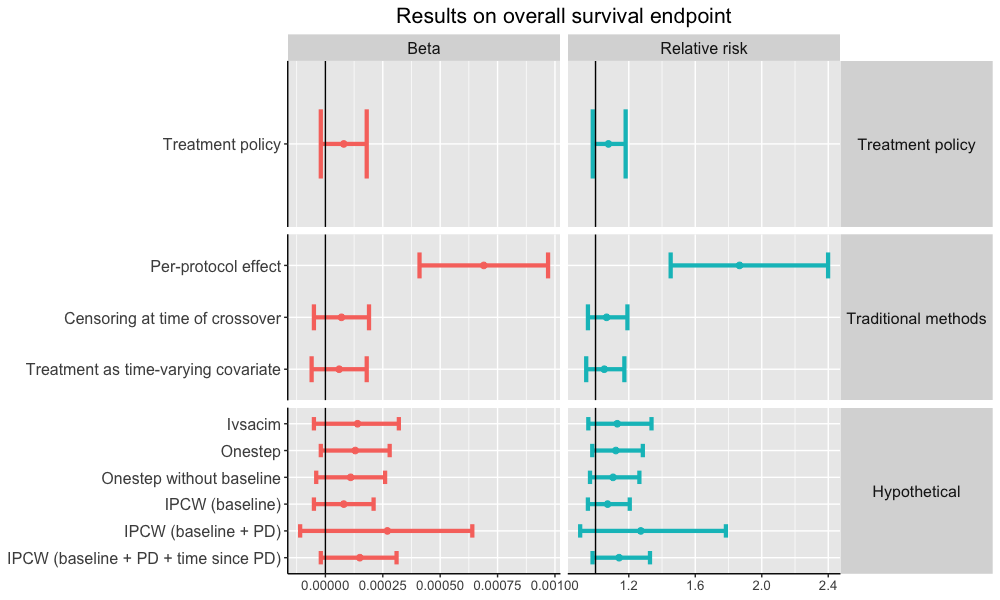}
	\caption{Results of the data analysis on the HELIOS trial, both for the additive effect on the hazard ($\beta$) and the relative risk. }
	\label{Fig:Data analysis results}
\end{figure}

In total, 41 patients (7\%) were censored during the trial, for unknown reasons. This censoring was not found to depend on the randomized arm, or on one of the baseline covariates. 
Table \ref{table: Data analysis results} and figure \ref{Fig:Data analysis results} summarize the results of the data analysis, while the obtained models and weights and predicted survival curves "if all patients would take control treatment and not cross over" are shown in Appendix \ref{Section: Appendix Data analysis Results }. The treatment policy relative risk (1.08, 95\% CI [0.98; 1.18]) shows an effect that is not significant. As expected, the hypothetical estimands, both using the \texttt{ivsacim} estimator (1.13, 95\% CI [0.96; 1.34]) and the one-step estimator (1.12, 95\% CI [0.98; 1.28]), indicate a somewhat larger treatment effect, as the setting is considered where control patients do not experience the possibly beneficial effect of crossover. However, the confidence intervals are a little wider too. In addition, the variance of the proposed one-step estimator ($7 \times 10^{-5}$) is smaller than the one of the \texttt{ivsacim} estimator ($9 \times 10^{-5}$).  The $p$-value obtained using the one-step estimator is 8\%, lower than the  $p$-value obtained by the \texttt{ivsacim} estimator (18\%). If no baseline covariates are included in the one-step estimator, the same $p$-value (12\%) as in the treatment policy analysis is obtained. In addition, we found the estimated function $\hat{E}(Z | \tilde{T}(0) \geq t, \bfL)$  not to depend on $t$, as expected if  $Z \independent \tilde{T}(0) | \bfL$ holds (see Appendix \ref{Section: Appendix Hypothetical estimand: Onestep estimator }). 
Moreover, all IV methods result in realistic counterfactual survival curves (see Appendix \ref{Section: Appendix Data analysis Results }), indicating that the control patients would have a somewhat worse survival curve if they were not allowed to cross over. 

The IPCW estimator using only baseline covariates as predictors (1.07, 95\% CI [0.95; 1.21]) does not perform better than simply censoring patients at the time of crossover, without correction for selection bias.  This indicates that some important confounders of the crossover-outcome relationship may not have been included. 
 When disease progression is included as a time-varying predictor in the inverse probability weights, we obtained a larger,   treatment effect (1.27, 95\% CI [0.91; 1.78])  than the treatment policy estimand. However, the model for the probability of crossover does not fit the data well and the standard error of the time-varying predictor for disease progression is very large (see Appendix \ref{Section: Appendix Data analysis Results Hypothetical estimand: IPCW estimators}). This results in a very wide confidence interval for the relative risk. In addition, the counterfactual survival curve is not realistic as the effect of crossover seems to be overrated (see Appendix \ref{Section: Appendix Data analysis Results }). However, no indications of positivity violations were  observed. When instead the time since disease progression is included as predictor, the model for the probability of crossover fits the data better and an effect (1.14, 95\% CI [0.98; 1.33]) in line with the one-step estimator is obtained.  However, the $p$-value of the IPCW estimator is smaller (6\%), which may be due to the fact that  additional model assumptions are being made.  Moreover, as shown by Latimer et al. \cite{latimer2014adjusting}, these IPCW estimates might be biased as a lot of patients crossed over in the HELIOS trial.

From the results, it is also clear that  the traditional methods, which are prone to selection bias when selective crossover occurs, should ideally be avoided.  In particular, the per-protocol effect, which excludes patients who cross over, gives an effect (1.87, 95\% CI [1.47; 2.40]) that is not at all in line with the other results (see figure \ref{Fig:Data analysis results}). In addition, a lot of data is removed as 63\% of the control patients cross over.  When treating treatment as a time-varying treatment the effect (1.05, 95\% CI [0.94; 1.17]) moves somewhat towards one, compared to the treatment policy estimand. This effect is very close to the effect where patients are censored at the time of crossover (1.07, 95\% CI [0.95; 1.19]).  

The  fact that the results of this data analysis on the overall survival endpoint are not significant is consistent with the published interim results of the HELIOS trial \cite{chanan2016ibrutinib}. 
However, the results presented in this paper  are intended for illustrative purposes only. 

	\begin{sidewaystable}
	\centering
	\begin{tabular}{  l l c c  c c c } 
		\toprule
			 \textbf{Estimand}   &  \textbf{Estimator} &  \textbf{$\mathbf{\beta}$}& \multicolumn{3}{c}{\textbf{Relative risk}} \\
		 & & \textbf{Estimate (SE)} & \textbf{Estimate} & \textbf{95\% CI} & \textbf{$p$-value}  \\
 \midrule
  \textbf{Treatment policy} &  Aalen least squares & $8 \times 10^{-5}$ ($5 \times 10^{-5}$) & 1.08 & [0.98; 1.18] & 0.12\\ 
    \textbf{Per-protocol effect} &  Aalen least squares& $69 \times 10^{-5}$ ($15 \times 10^{-5}$)  & 1.87 & [1.45; 2.40] & 0.00 \\ 
 (excluding patients \\ 
 who cross over) \\
 \textbf{Censoring patients} &  Aalen least squares & $7 \times 10^{-5}$ ($6 \times 10^{-5}$)  & 1.07  & [0.95; 1.19] & 0.25 \\
 \textbf{at time of crossover} \\
 \textbf{As treated}  &  Aalen least squares &  $6 \times 10^{-5}$ ($6 \times 10^{-5}$)&1.05 & [0.94; 1.17] & 0.35  \\
 (Treatment as time- \\
 varying covariate) \\
\textbf{Hypothetical } 
& IV estimators  \\  
& \hspace*{15pt} Ivsacim \cite{ying2022ivsacim}  &  $14 \times 10^{-5}$ ($9 \times 10^{-5}$)  &1.13  & [0.96; 1.34] & 0.18\\  
& \hspace*{15pt} One-step estimator & $13 \times 10^{-5}$ ($7 \times 10^{-5}$)  &1.12 & [0.98; 1.28] & 0.08 \\  
& \hspace*{15pt} One-step estimator & $11 \times 10^{-5}$ ($7 \times 10^{-5}$)  &1.11 & [0.97; 1.26] & 0.12 \\  
 &   \hspace*{20pt} without baseline covariates \\
& IPCW estimators \\
& \hspace*{15pt} Baseline covariates as predictors &$8 \times 10^{-5}$ ($6 \times 10^{-5}$)  &1.07  & [0.95; 1.21] & 0.22  \\
 & \hspace*{15pt} Baseline covariates and &    $27 \times 10^{-5}$ ($19 \times 10^{-5}$)  &1.27  & [0.91; 1.78] & 0.15 \\ 
   & \hspace*{20pt} indicator for PD as predictors \\
   & \hspace*{15pt} Baseline covariates, indicator for   & $15 \times 10^{-5}$ ($8 \times 10^{-5}$)  &1.14  & [0.98; 1.33] & 0.06  \\
   &   \hspace*{20pt} and time since PD as predictors \\
	\bottomrule
	\end{tabular}
	\caption{Results of the data analysis on the HELIOS trial. \\ PD: progressive disease}
	\label{table: Data analysis results}
	\end{sidewaystable}

\section{Discussion}

In this article, we considered the setting in randomized clinical trials with time-to-event endpoints, where patients can cross over from the control to the experimental arm, typically after disease progression.  We explained how traditional methods to correct for crossover easily lure one into making large extrapolations. In view of this, 
we discussed a structural nested cumulative survival time model that can be used to target the hypothetical estimand "if all patients would stay on the assigned treatment arm", summarized as a survival risk ratio. An estimator for this effect, developed by Ying and Tchetgen Tchetgen \cite{ying2022ivsacim}, was compared to the one-step estimator we developed. These estimators do not suffer from positivity or unconfoundedness assumptions, as may often be the case with IPCW methods. In particular, these methods use randomization as an instrumental variable. 
The developed one-step estimator, in contrast to the  estimator by Ying and Tchetgen Tchetgen \cite{ying2022ivsacim}, is a doubly robust estimator, meaning that it is unbiased even if one of the working models is misspecified.  In addition, data-adaptive methods (e.g. variable selection) can be used for the model for the hazard of death. 

This paper also aimed to give more insight into the model proposed by Ying and Tchetgen Tchetgen \cite{ying2022ivsacim}, and apply it to a real case study. Moreover, we aimed to 
obtain insight into the performance of the instrumental variable estimator in settings with a lot of crossover, as  in the data analysis performed by Ying and Tchetgen Tchetgen \cite{ying2022ivsacim} the number of switchers was rather limited, i.e. 5\% of the treated patients and 10\% of the control patients crossed over. 
All instrumental variable estimators showed good performance in the conducted data analysis, even though 63\% of the control patients switched treatment. 

In this paper, we only considered the setting where patients can cross over from one arm to the other.  However, SNCSTM (\ref{eq: SNCSTM general}) can be extended to the setting where patients can cross over from both arms: 
\begin{align}
	\frac{P[\tilde{T}(\overline{D}(t_m), 0)>t  | \overline{D}(t_m), Z, \bfL, \tilde{T}\geq t_m]}{ P[\tilde{T}(\overline D(t_{m-1}), 0)>t  | \overline{D}(t_m), Z, \bfL, \tilde{T}\geq t_m]}  
	= \exp\left\{-\beta D(t_m) (t\wedge t_{m+1} - t_m) \right\}, 
	\label{eq: SNCSTM both arms}
\end{align}
with $\tilde{T}(\overline D(t_m), 0)$ the potential time to event had the patient, possibly contrary to fact,  taken treatments $\overline D(t_m)$ up to time $t_m$ and control treatment thereafter. However, in model (\ref{eq: SNCSTM both arms}) it is assumed that the ratio of survival probabilities does not depend on $Z$, which might be a strong assumption.

\bibliographystyle{abbrv}
\bibliography{mybib}

\newpage 

\begin{appendices}
	
	\renewcommand\thefigure{\arabic{figure}}    
	\setcounter{figure}{0} 
	\renewcommand\thetable{\arabic{table}}    
	\setcounter{table}{0}

%

\section{SNCSTM}
\label{Section: Appendix SNCSTM}

In this appendix, we discuss the SNCSTM proposed by Ying and Tchetgen Tchetgen \cite{yingstructural}.    Let $\tilde{T}(\overline d(t_m), 0)$ denote the potential time to event had the patient, possibly contrary to fact,  taken treatments $\overline d(t_m)$ up to time $t_m$ and control treatment thereafter.

In the SNCSTM, one contrasts the ratio of survival probabilities at time $t$ upon starting experimental treatment at time  $t_m \leq t$ versus at time $t_{m+1}$, for patients still alive at time $t_m$,  in arm $Z$, with treatment history $\overline D(t_m)$ and with baseline covariates $\bfL$:
\begin{align}
	&\frac{P[\tilde{T}(\overline{D}(t_m), 0)>t  | \overline{D}(t_m), Z, \bfL, \tilde{T}\geq t_m]}{ P[\tilde{T}(\overline D(t_{m-1}), 0)>t  | \overline{D}(t_m), Z, \bfL, \tilde{T}\geq t_m]} \nn \\ &= \exp\left\{- \int_{t_m}^{t\wedge t_{m+1}} D(t_m) dB(s)\right\} \nn \\
	&= \exp\left\{- D(t_m) (B(t\wedge t_{m+1}) - B(t_m)) \right\}, 
	\label{eq: appendix SNCSTM general}
\end{align}
for an unknown function $B(.)$. 
In this model, $dB(t_m)$ can be interpreted  as a measure of the treatment effect for patients who share the same history and take control at time $t_m$  and are treated thereafter, compared to those who are treated after time $t_{m-1}$. 
A positive value implies that the treatment increases patients' survival probability. If $D(t_m) = 0$, function (\ref{eq: appendix SNCSTM general}) becomes 1, because then the two treatment regimes being compared are identical.

In the main paper, model (\ref{eq: SNCSTM general}), is referred to as the  `constant hazards difference model' \cite{yingstructural}, corresponding to the  choice  
\begin{equation*}
	B(t) \equiv \beta t, 
	\label{eq: SNCSTM time constant}
\end{equation*}
for a constant $\beta$ and conditioning on the control arm $Z=1$ as in our setting patients can only cross over from the control to the experimental arm and not vice versa. 


\newpage 

\section{Hypothetical estimand}
\label{Section: Appendix Hypothetical estimand}

In this appendix, we show that under model (\ref{eq: appendix SNCSTM general}), and the no-current treatment value interaction assumption (see below), it holds that 
\begin{equation}
	\frac{P[\tilde{T}(0)>t |  \bfL ]}{P[\tilde{T}(1)>t |  \bfL]} =  \exp\left\{\int_{0}^{t}  dB(s)\right\} = \exp\left\{B(t)\right\}. 
	\label{eq: appendix hypothetical survival risk ratio}
\end{equation}
The no-current treatment value interaction assumption \cite{robins1994correcting,yingstructural} states that the causal effect of control  at time $t_m$ versus experimental treatment among patients who were taking control at time $t_m$ is equal to that among patients who were treated at time $t_m$ conditional on past history \cite{yingstructural}: 
\begin{eqnarray}
 	&&\frac{P[\tilde{T}(\overline{D}(t_{m-1}),D(t_m)=1,0)>t  | \overline{D}(t_{m-1}),D(t_m)=1, Z, \bfL, \tilde{T}\geq t_m]}{P[\tilde{T}(\overline{D}(t_{m-1}),0)>t  | \overline{D}(t_{m-1}),D(t_m)=1, Z, \bfL, \tilde{T}\geq t_m]} \nn \\
 	&&= 	\frac{P[\tilde{T}(\overline{D}(t_{m-1}),D(t_m)=1,0)>t  | \overline{D}(t_{m-1}),D(t_m)=0, Z, \bfL, \tilde{T}\geq t_m]}{P[\tilde{T}(\overline{D}(t_{m-1}),0)>t  | \overline{D}(t_{m-1}),D(t_m)=0, Z, \bfL, \tilde{T}\geq t_m]}.   \label{eq: appendix ass no interaction general}  
\end{eqnarray}
In the setting where patients can only cross over from the control to the experimental arm and not vice versa, this assumption boils down to 
\begin{eqnarray}
	&&\frac{P[\tilde{T}(\overline{D}(t_{m})=1,0)>t  | \overline{D}(t_{m})=1, Z=1, \bfL, \tilde{T}\geq t_m]}{P[\tilde{T}(\overline{D}(t_{m-1})=1,0)>t  | \overline{D}(t_m)=1, Z=1, \bfL, \tilde{T}\geq t_m]} \nn \\
	&&= 	\frac{P[\tilde{T}(\overline{D}(t_{m})=1,0)>t  | \overline{D}(t_{m-1})=1,D(t_m)=0, Z=1, \bfL, \tilde{T}\geq t_m]}{P[\tilde{T}(\overline{D}(t_{m-1})=1,0)>t  | \overline{D}(t_{m-1})=1,D(t_m)=0, Z=1, \bfL, \tilde{T}\geq t_m]},   \label{eq: appendix ass no interaction general control}  
\end{eqnarray}
stating that the causal effect  of switching at time $t_{m+1}$ versus at time  $t_{m}$ is the same for control patients who in fact did switch at time $t_m$ and for those who did not, with the same baseline covariates. This assumption could be violated if patients who cross over do not experience the same treatment effect as those who do not. This assumption can be relaxed by including time-varying covariates in the SNCSTM, but this will complicate the estimation procedure.

\noindent \textbf{Proof of (\ref{eq: appendix hypothetical survival risk ratio})}
 
For $t \in [t_1,t_2[$, model (\ref{eq: appendix SNCSTM general}) implies 
\begin{equation*}
	\frac{P[\tilde{T}(1)>t  | {D}(t_1) = 1, Z, \bfL, \tilde{T}\geq t_1]}{P[\tilde{T}(0)>t  | {D}(t_1) = 1, Z, \bfL, \tilde{T}\geq t_1]} =  \exp\left\{- \int_{t_1}^{t}  dB(s)\right\} 
\end{equation*}
and assumption (\ref{eq: appendix ass no interaction general}) implies
\begin{equation*}
	\frac{P[\tilde{T}(1)>t  | {D}(t_1) = 1, Z, \bfL, \tilde{T}\geq t_1]}{P[\tilde{T}(0)>t  | {D}(t_1) = 1, Z, \bfL, \tilde{T}\geq t_1]} = 	\frac{P[\tilde{T}(1)>t  | {D}(t_1) = 0, Z, \bfL, \tilde{T}\geq t_1]}{P[\tilde{T}(0)>t  | {D}(t_1) = 0, Z, \bfL, \tilde{T}\geq t_1]}. 
\end{equation*}
It follows that, 
\begin{eqnarray*}
  &&P(\tilde{T}(0)>t  | Z, \bfL) \\
  &&=  P(\tilde{T}(0)>t  |  {D}(t_1) = 1,Z, \bfL) P( {D}(t_1) = 1 |  Z,\bfL ) + P(\tilde{T}(0)>t  |  {D}(t_1) = 0,Z, \bfL) P( {D}(t_1) = 0 |  Z,\bfL ) \\ 
  &&= P(\tilde{T}(0)>t  |  {D}(t_1) = 1,Z, \bfL,\tilde{T}\geq t_1)P(\tilde{T}\geq t_1  |  {D}(t_1) = 1,Z,\bfL) P( {D}(t_1) = 1 |  Z,\bfL ) \\
  &&+ P(\tilde{T}(0)>t  |  {D}(t_1) = 0,Z, \bfL,\tilde{T}\geq t_1)P(\tilde{T}\geq t_1  |  {D}(t_1) = 0,Z,\bfL) P( {D}(t_1) = 0 |  Z,\bfL )  \\
  && = \exp\left\{\int_{t_1}^{t}  dB(s)\right\} \left[P(\tilde{T}(1)>t  |  {D}(t_1) = 1,Z, \bfL,\tilde{T}\geq t_1)P(\tilde{T}\geq t_1  |  {D}(t_1) = 1,Z,\bfL) P( {D}(t_1) = 1 |  Z,\bfL )  \right. \\
  && \left. +   P(\tilde{T}(1)>t  |  {D}(t_1) = 0,Z, \bfL,\tilde{T}\geq t_1)P(\tilde{T}\geq t_1  |  {D}(t_1) = 0,Z,\bfL) P( {D}(t_1) = 0 |  Z,\bfL )  \right] \\ 
  &&=\exp\left\{\int_{t_1}^{t}  dB(s)\right\}  P(\tilde{T}(1)>t  | Z, \bfL). 
\end{eqnarray*}
Assuming $Z \independent \tilde{T}(0) | \bfL$  and $Z \independent \tilde{T}(1) | \bfL$, it follows that 
\begin{equation*}
P(\tilde{T}(1)>t  | \bfL)  = \exp\left\{- \int_{t_1}^{t}  dB(s)\right\} P(\tilde{T}(0)>t  |  \bfL).
\end{equation*}

For $t \in [t_2,t_3[$, model (\ref{eq: appendix SNCSTM general}) implies 
\begin{align}
	\frac{P[\tilde{T}(D(t_1),1)>t  | D(t_1), {D}(t_2) = 1, Z, \bfL, \tilde{T}\geq t_2]}{P[\tilde{T}(D(t_1),0)>t  | D(t_1), {D}(t_2) = 1, Z, \bfL, \tilde{T}\geq t_2]} &=  \exp\left\{- \int_{t_2}^{t}  dB(s)\right\} \label{eq: appendix SNCSTM M2 eq1}  \\ 
		\frac{P[\tilde{T}(1,0)>t  |{D}(t_1) = 1, Z, \bfL, \tilde{T}\geq t_1]}{P[\tilde{T}(0,0)>t  |{D}(t_1) = 1, Z, \bfL, \tilde{T}\geq t_1]} &=  \exp\left\{- \int_{t_1}^{t_2}  dB(s)\right\}  \label{eq: appendix SNCSTM M2 eq2}
\end{align}
and assumption (\ref{eq: appendix ass no interaction general}) implies
\begin{align}
\frac{P[\tilde{T}(D(t_1),1)>t  | D(t_1), {D}(t_2) = 1, Z, \bfL, \tilde{T}\geq t_2]}{P[\tilde{T}(D(t_1),0)>t  | D(t_1), {D}(t_2) = 1, Z, \bfL, \tilde{T}\geq t_2]} &= 	\frac{P[\tilde{T}(D(t_1),1)>t  | D(t_1), {D}(t_2) = 0, Z, \bfL, \tilde{T}\geq t_2]}{P[\tilde{T}(D(t_1),0)>t  | D(t_1), {D}(t_2) = 0, Z, \bfL, \tilde{T}\geq t_2]}   \label{eq: appendix ass no interaction M2 eq1}  \\
		\frac{P[\tilde{T}(1,0)>t  | {D}(t_1) = 1, Z, \bfL, \tilde{T}\geq t_1]}{P[\tilde{T}(0,0)>t  | {D}(t_1) = 1, Z, \bfL, \tilde{T}\geq t_1]}  &= 	\frac{P[\tilde{T}(1,0)>t  | {D}(t_1) = 0, Z, \bfL, \tilde{T}\geq t_1]}{P[\tilde{T}(0,0)>t  | {D}(t_1) = 0, Z, \bfL, \tilde{T}\geq t_1]}. \label{eq: appendix ass no interaction M2 eq2}
\end{align}
From model (\ref{eq: appendix SNCSTM M2 eq1}) and assumption (\ref{eq: appendix ass no interaction M2 eq1}) it follows that 
\begin{eqnarray}
&&P(\tilde{T}(D(t_1),0)>t  | Z, \bfL) \nn \\
&&= E\left[P(\tilde{T}(D(t_1),0)>t  | Z, \bfL,D(t_1))  | Z, \bfL \right] \nn\\
&&= E\left[   \sum_{d_2 = 0}^{1} P(\tilde{T}(D(t_1),0)>t  | Z, \bfL,D(t_1),D(t_2) = d_2,\tilde{T}(D(t_1),0)\geq t_2  ) \right. \nn \\
&& \left. \times  P(\tilde{T}(D(t_1),0)\geq t_2  | Z, \bfL,D(t_1),D(t_2) = d_2 )  P(D(t_2) = d_2 | Z, \bfL,D(t_1))| Z, \bfL \right] \nn \\
&&=  E\left[   \sum_{d_2 = 0}^{1} P(\tilde{T}(D(t_1),0)>t  | Z, \bfL,D(t_1),D(t_2) = d_2,\tilde{T}\geq t_2  ) \right. \nn \\
&& \left. \times  P(\tilde{T}\geq t_2  | Z, \bfL,D(t_1),D(t_2) = d_2 )  P(D(t_2) = d_2 | Z, \bfL,D(t_1))| Z, \bfL \right]\nn \\
&&=  E\left[   \sum_{d_2 = 0}^{1} \exp\left\{\int_{t_2}^t dB(s)\right\} P(\tilde{T}(D(t_1),1)>t  | Z, \bfL,D(t_1),D(t_2) = d_2,\tilde{T}\geq t_2  ) \right.\nn \\
&& \left. \times  P(\tilde{T}\geq t_2  | Z, \bfL,D(t_1),D(t_2) = d_2 )  P(D(t_2) = d_2 | Z, \bfL,D(t_1))| Z, \bfL \right] \nn\\
&&=  \exp\left\{\int_{t_2}^t dB(s)\right\}  E\left[    P(\tilde{T}(D(t_1),1)>t  | Z, \bfL,D(t_1)  | Z, \bfL \right] \nn\\
&&=   \exp\left\{\int_{t_2}^t dB(s)\right\}  P(\tilde{T}(D(t_1),1)>t  | Z, \bfL), 
\label{eq: appendix model M2}
\end{eqnarray}
where the third equality follows from the consistency assumption.
Model (\ref{eq: appendix SNCSTM M2 eq2}) and assumption (\ref{eq: appendix ass no interaction M2 eq2})
imply 
\begin{eqnarray*}
	&&P(\tilde{T}(0,0)>t  | Z, \bfL) \nn \\
	&&= \sum_{d_1 = 0}^{1}  P(\tilde{T}(0,0)>t  | Z, \bfL, D(t_1) = d_1, \tilde{T}\geq t_1 ) P(\tilde{T}\geq t_1 | Z, \bfL, D(t_1) = d_1 ) P(D(t_1) = d_1, | Z, \bfL  )  \\ 
	&&= \exp\left\{\int_{t_1}^{t_2} dB(s)\right\} P(\tilde{T}(1,0)>t  | Z, \bfL, D(t_1) = d_1, \tilde{T}\geq t_1 ) \\ && \times P(\tilde{T}\geq t_1 | Z, \bfL, D(t_1) = d_1 ) P(D(t_1) = d_1, | Z, \bfL  )  \\ 
	&&= \exp\left\{\int_{t_1}^{t_2} dB(s)\right\}  P(\tilde{T}(1,0)>t  | Z, \bfL). 
\end{eqnarray*}
From (\ref{eq: appendix model M2}) it then follows that 
\begin{eqnarray*}
P(\tilde{T}(0,0)>t  | Z, \bfL) 
	&&= \exp\left\{\int_{t_1}^{t_2} dB(s)\right\}  \exp\left\{\int_{t_2}^{t} dB(s)\right\} P(\tilde{T}(1,1)>t  | Z, \bfL)  \\
	&&= \exp\left\{\int_{t_1}^{t_t} dB(s)\right\}  P(\tilde{T}(1,1)>t  | Z, \bfL). 
\end{eqnarray*}
Assuming $Z \independent \tilde{T}(0) | \bfL$  and $Z \independent \tilde{T}(0) | \bfL$, it follows that 
\begin{equation*}
	P(\tilde{T}(1,1)>t  | \bfL)  = \exp\left\{- \int_{t_1}^{t}  dB(s)\right\} P(\tilde{T}(0,0)>t  |  \bfL).
\end{equation*}
For general $t$,  it can likewise be shown by induction that 
\begin{equation*}
	P(\tilde{T}(1)>t  | \bfL) = \exp\left\{- \int_{t_1}^{t}  dB(s)\right\}P(\tilde{T}(0)>t | \bfL ) . 
\end{equation*}
\qedwhite

\newpage
\section{One-step estimator}
\label{Section: Appendix one-step estimator}

\subsection{Proof of unbiasedness of estimating equation}
\label{Section: Appendix proof of estimating equation}

\begin{theorem}
	\begin{equation*}
		\small
	E \left[ \left\{Z-E(Z | \tilde{T}(0) \geq t, \bfL   ) \right\} \exp\left\{ \int_{0}^{t}D(s)dB(s) \right\}  \left\{dN(t) -Y(t)D(t)dB(s) -Y(t)\lambda^0(t |  \bfL)dt \right\} \right] = 0, 
\end{equation*}
with $\lambda^0(t |  \bfL)$ the hazard rate function of $\tilde{T}(0) $ conditional on  $\bfL$
 and 
\begin{equation*}
	E(Z | \tilde{T}(0) \geq t, \bfL) = \frac{E\left[Z Y(t)\exp\left\{\int_{0}^{t}D(s)dB(s) \right\} |\bfL  \right]}{E\left[Y(t)\exp\left\{ \int_{0}^{t}D(s)dB(s) \right\} |\bfL  \right]}. 
\end{equation*}
assuming $Z \independent \tilde{T}(0) | \bfL$  and $C\independent (\tilde{T},Z, \overline{D}(t)) |\bfL $.
\label{theorem: Estimating equation}
\end{theorem}
To prove this, we first need some lemmas. 
\begin{lem}
If SNCSTM (\ref{eq: appendix SNCSTM general}) holds, and assuming $Z \independent \tilde{T}(0) | \bfL$ and $C\independent (\tilde{T},Z, \overline{D}(t)) |\bfL $, \\ it follows that 
\begin{eqnarray}
P(\tilde{T}(0) \geq t |Z, \bfL) &&=  E\left(I(\tilde{T}\geq t) \exp\left\{\int_{0}^{t}D(s)dB(s) \right\} |Z, \bfL \right)  \label{eq: survival T0 counterfactual} \\   
P(\tilde{T}(0) \geq t |Z, \bfL) &&=  E\left(\frac{Y(t)}{P(C\geq t| \bfL)}\exp\left\{\int_{0}^{t}D(s)dB(s) \right\} |Z, \bfL \right)  \label{eq: survival T0 conditional} \\ 
P(\tilde{T}(0) \geq t |\bfL) &&=  E\left( \frac{Y(t)}{P(C\geq t| \bfL)}\exp\left\{\int_{0}^{t}D(s)dB(s) \right\} | \bfL \right).   \label{eq: survival T0} \\
P(\tilde{T}(0) \geq t) &&=  E\left( \frac{Y(t)}{P(C\geq t| \bfL)}\exp\left\{\int_{0}^{t}D(s)dB(s) \right\}  \right).   \label{eq: survival T0 marginal}
\end{eqnarray}
\label{Lemma: counterfactual survival probability}
\end{lem}

\noindent \textbf{Proof}

\noindent 
	For $t \in [t_1,t_2[$, we have that: 
\begin{eqnarray*}
	P(\tilde{T}(0) \geq t |Z, \bfL) &&= E\left[I\left(\tilde{T}(0) \geq t\right) |Z, \bfL \right] \\
	&&=  E\left[ E\left( I\left(\tilde{T}(0) \geq t\right) |D(t_1),Z, \bfL  \right) |Z, \bfL \right] \\
	&&=  E\left[ P(\tilde{T}(D(t_1)) \geq t  |D(t_1),Z, \bfL) \exp\left\{D(t_1)(B(t)-B(t_1))  \right\}  |Z, \bfL \right] \\
	&&=  E\left[ P(\tilde{T} \geq t  |D(t_1),Z, \bfL) \exp\left\{D(t_1)(B(t)-B(t_1))  \right\}  |Z, \bfL \right] \\
	&&=  E\left[ E\left( I(\tilde{T} \geq t)  |D(t_1),Z, \bfL\right) \exp\left\{D(t_1)(B(t)-B(t_1))  \right\}  |Z, \bfL \right] \\
	&&= E\left[  I(\tilde{T}\geq t)  \exp\left\{D(t_1)(B(t)-B(t_1))  \right\}  |Z, \bfL \right], 
\end{eqnarray*}
where the third equation follows from SNCSTM	(\ref{eq: appendix SNCSTM general})  and the fourth from the consistency assumption. 

For $t \in [t_2,t_3[$, we have that: 
	\begin{eqnarray*}
	P(\tilde{T}(0) \geq t |Z, \bfL)  	&&=  E\left[ E\left( I\left(\tilde{T}(0) \geq t\right) |D(t_1),Z, \bfL  \right) |Z, \bfL \right] \\
	&&=  E\left[ P(\tilde{T}(D(t_1),0) \geq t  |D(t_1),Z, \bfL) \exp\left\{D(t_1)(B(t_2)-B(t_1))  \right\}  |Z, \bfL \right], 
\end{eqnarray*}
where the second equation follows from SNCSTM	(\ref{eq: appendix SNCSTM general}). This can be further rewritten as 
	\begin{eqnarray*}
&&	P(\tilde{T}(0) \geq t |Z, \bfL)  	\\
&&=  E\left[ E(E( I(\tilde{T}(D(t_1),0) \geq t ) |\overline{D}(t_2),Z, \bfL )   |D(t_1),Z, \bfL) \exp\left\{D(t_1)(B(t_2)-B(t_1))  \right\}  |Z, \bfL \right] \\
&&=  E\left[ E\left( P(\tilde{T}(D(t_1),0) \geq t   | \overline{D}(t_2),Z, \bfL, \tilde{T}(D(t_1),0) \geq t _2 )  \right. \right. \\
&& \left. \left.\times  P( \tilde{T}(D(t_1),0) \geq t _2| \overline{D}(t_2),Z, \bfL)  |D(t_1),Z, \bfL\right) \exp\left\{D(t_1)(B(t_2)-B(t_1))  \right\}  |Z, \bfL \right] \\
&&=  E\left[ E\left( P(\tilde{T}(D(t_1),0) \geq t   | \overline{D}(t_2),Z, \bfL, \tilde{T} \geq t _2 )  \right. \right. \\
&& \left. \left.\times  P( \tilde{T}(D(t_1),0) \geq t _2| \overline{D}(t_2),Z, \bfL)  |D(t_1),Z, \bfL\right) \exp\left\{D(t_1)(B(t_2)-B(t_1))  \right\}  |Z, \bfL \right] \\ 
&&=  E\left[ E\left( P(\tilde{T}(\overline{D}(t_2),0) \geq t   | \overline{D}(t_2),Z, \bfL, \tilde{T} \geq t _2 )  \exp\left\{D(t_2)(B(t)-B(t_2))  \right\}  \right. \right. \\
&& \left. \left.\times  P( \tilde{T}(D(t_1),0) \geq t _2| \overline{D}(t_2),Z, \bfL)  |D(t_1),Z, \bfL\right) \exp\left\{D(t_1)(B(t_2)-B(t_1))  \right\}  |Z, \bfL \right], 
\end{eqnarray*}
where the third equality follows from the consistency assumption and the fourth equality from SNCSTM	(\ref{eq: appendix SNCSTM general}).  Again using the consistency assumption, this can be further rewritten as
	\begin{eqnarray*}
	&&	P(\tilde{T}(0) \geq t |Z, \bfL)  	\\
	&&=  E\left[ E\left( P(\tilde{T}(\overline{D}(t_2),0) \geq t   | \overline{D}(t_2),Z, \bfL, \tilde{T}({D}(t_1),0)  \geq t _2 )  \exp\left\{D(t_2)(B(t)-B(t_2))  \right\}  \right. \right. \\
	&& \left. \left.\times  P( \tilde{T}(D(t_1),0) \geq t _2| \overline{D}(t_2),Z, \bfL)  |D(t_1),Z, \bfL\right) \exp\left\{D(t_1)(B(t_2)-B(t_1))  \right\}  |Z, \bfL \right] \\
	&&=  E\left[ E\left( P(\tilde{T} \geq t   | \overline{D}(t_2),Z, \bfL )  \exp\left\{D(t_2)(B(t)-B(t_2))  \right\}   |D(t_1),Z, \bfL\right)  \right.\\
	&& \left. \times   \exp\left\{D(t_1)(B(t_2)-B(t_1))  \right\}  |Z, \bfL \right] \\
	&&=  E\left[I(\tilde{T}\geq t) \exp\left\{\int_{t_1}^t D(s) dB(s)  \right\}     |Z, \bfL \right]. 
\end{eqnarray*}
For general $t$,  it can likewise be shown by induction that  
\begin{equation*}
P(\tilde{T}(0) \geq t |Z, \bfL) =  E\left(I(\tilde{T}\geq t)\exp\left\{\int_{0}^{t}D(s)dB(s) \right\} |Z, \bfL \right). 
\end{equation*}
From  $C\independent (\tilde{T},Z, \overline{D}(t)) |\bfL $ it follows that: 
\begin{eqnarray*}
 &&P(\tilde{T}(0) \geq t |Z, \bfL) \\ 
 &&=  E\left(I(\tilde{T}\geq t)\exp\left\{\int_{0}^{t}D(s)dB(s) \right\} |Z, \bfL \right) \\ 
 &&=  E\left(E\left(I(\tilde{T}\geq t)\exp\left\{\int_{0}^{t}D(s)dB(s) \right\} |Z, \bfL, \overline{D}(t) \right) |Z, \bfL \right) \\ 
 &&=  E\left(\frac{E(I(C\geq t) |\bfL)}{P(C\geq t| \bfL)} E\left(I(\tilde{T}\geq t)\exp\left\{\int_{0}^{t}D(s)dB(s) \right\} |Z, \bfL, \overline{D}(t) \right) |Z, \bfL \right) \\ 
 &&=  E\left(\frac{E(I(C\geq t) |Z,\bfL, \overline{D}(t) )}{P(C\geq t| \bfL)} E\left(I(\tilde{T}\geq t)\exp\left\{\int_{0}^{t}D(s)dB(s) \right\} |Z, \bfL, \overline{D}(t) \right) |Z, \bfL \right) \\ 
  &&=  E\left(\frac{1}{P(C\geq t| \bfL)} E\left(I(C\geq t) I(\tilde{T}\geq t)\exp\left\{\int_{0}^{t}D(s)dB(s) \right\} |Z, \bfL, \overline{D}(t) \right) |Z, \bfL \right) \\ 
 &&=  E\left(\frac{I(C\geq t) I(\tilde{T}\geq t)}{P(C\geq t| \bfL)}  \exp\left\{\int_{0}^{t}D(s)dB(s) \right\} |Z, \bfL \right) \\ 
 &&=  E\left(\frac{Y(t)}{P(C\geq t| \bfL)}\exp\left\{\int_{0}^{t}D(s)dB(s) \right\} |Z, \bfL \right).
\end{eqnarray*}
In addition, from $Z \independent \tilde{T}(0) | \bfL$ it follows that
\begin{equation*}
	P(\tilde{T}(0) \geq t | \bfL) =  E\left(\frac{Y(t)}{P(C\geq t| \bfL)}\exp\left\{\int_{0}^{t}D(s)dB(s) \right\} | \bfL \right). 
\end{equation*}
Consequently, 
\begin{equation*}
	P(\tilde{T}(0) \geq t) =  E\left(\frac{Y(t)}{P(C\geq t| \bfL)}\exp\left\{\int_{0}^{t}D(s)dB(s) \right\} \right). 
\end{equation*}
Therefore, the counterfactual survival probabilities `when always taking control treatment'  can  be obtained by removing the effect of non-zero treatment at each period from the beginning of the study period to the end. 
	\qedwhite
	

\begin{lem}
	If SNCSTM (\ref{eq: SNCSTM general}) holds, and assuming $Z \independent \tilde{T}(0) | \bfL$ and $C\independent (\tilde{T},Z, \overline{D}(t)) |\bfL $, \\ it follows that 
\begin{equation*}
		E(Z | \tilde{T}(0) \geq t, \bfL)	= \frac{E\left(ZY(t) \exp\left\{\int_{0}^{t}D(s)dB(s) \right\}  |\bfL \right)  }{E\left(Y(t) \exp\left\{\int_{0}^{t}D(s)dB(s) \right\} |\bfL  \right)}.
\end{equation*}
\end{lem}
\textbf{Proof}

\noindent From Lemma \ref{Lemma: counterfactual survival probability} it follows that 
	\begin{eqnarray*}
		E(Z | \tilde{T}(0) \geq t, \bfL) &&= \int Z f(Z| \tilde{T}(0) \geq t, \bfL)dZ \\
		&&=  \int Z \frac{	P(\tilde{T}(0) \geq t |Z, \bfL) }{P(\tilde{T}(0) \geq t |\bfL) } f(Z| \bfL) dZ \\
		&&= \int Z \frac{	E\left( Y(t)\exp\left\{\int_{0}^{t}D(s)dB(s) \right\} |Z, \bfL \right)  }{E\left( Y(t)\exp\left\{\int_{0}^{t}D(s)dB(s) \right\} | \bfL \right)  } f(Z| \bfL) dZ. 
	\end{eqnarray*}
Consequently, 
	\begin{eqnarray*}
	E(Z |  \tilde{T}(0) \geq t, \bfL)	&&= \frac{E\left(ZY(t) \exp\left\{\int_{0}^{t}D(s)dB(s) \right\}  |\bfL \right)  }{E\left(Y(t) \exp\left\{\int_{0}^{t}D(s)dB(s) \right\} |\bfL  \right)}.
\end{eqnarray*}
\qedwhite

\noindent \textbf{Proof of Theorem \ref{theorem: Estimating equation}}	

\noindent From equation (\ref{eq: survival T0 counterfactual}) and the assumption $Z \independent \tilde{T}(0) | \bfL$, it follows that 
\begin{equation*}
	P(\tilde{T}(0) \geq t |\bfL) =  E\left(\tilde{Y}(t)\exp\left\{\int_{0}^{t}D(s)dB(s) \right\} |Z, \bfL \right), 
\end{equation*}
with $\tilde{Y}(t) =  I(\tilde{T}\geq t)$.
By taking logs of each side of this equation and differentiating with respect to $t$, it follows that 
\begin{eqnarray*}
	\lambda^0(t |  \bfL) 
	&&= - \frac{d}{dt} \log E \left[\tilde{Y}(t)  \exp\left\{\int_{0}^{t}D(s)dB(s) \right\}  |  Z, \bfL \right], 
\end{eqnarray*}
with $\lambda^0(t |  \bfL)$ the hazard rate function of $\tilde{T}(0) $ conditional on  $\bfL$.

Let $\tilde{S}(t | \overline{D}(t),Z, \bfL)$ denote the survival probability at time $t$ conditional on exposures $\overline{D}(t)$, treatment assignment $Z$ and baseline covariates $\bfL$ and $\tilde{f}(t | \overline{D}(t), Z, \bfL)$ the density function of $\tilde{T}$ at time $t$ conditional on exposures $\overline{D}(t)$, treatment assignment $Z$ and baseline covariates $\bfL$.  It then follows that
\begin{eqnarray*}
	&&\lambda^0(t |  \bfL) \\
	&&= - \frac{d}{dt} \log E \left[\tilde{S}(t | \overline{D}(t),Z, \bfL ) \exp\left\{\int_{0}^{t}D(s)dB(s) \right\} |  Z, \bfL \right]\\ 
	&&= - \frac{ E\left[-\tilde{f}(t |\overline{D}(t),Z, \bfL ) \exp\left\{\int_{0}^{t}D(s)dB(s) \right\}+\tilde{S}(t | \overline{D}(t),Z, \bfL ) \frac{d}{dt} \left(   \exp\left\{\int_{0}^{t}D(s)dB(s) \right\} \right) |  Z, \bfL  \right]}{E \left[\tilde{S}(t | \overline{D}(t),Z, \bfL) \exp\left\{\int_{0}^{t}D(s)dB(s) \right\} |  Z, \bfL \right]}.
\end{eqnarray*}
It follows that 
\begin{eqnarray*}
	\lambda^0(t |  \bfL) 
	&&=  \frac{ E\left[ \tilde{S}(t |\overline{D}(t),Z, \bfL) \exp\left\{\int_{0}^{t}D(s)dB(s) \right\}  \left\{\frac{\tilde{f}(t | \overline{D}(t),Z, \bfL )}{\tilde{S}(t | \overline{D}(t),Z, \bfL )} - D(t)\frac{dB(t)}{dt} \right\} |  Z, \bfL   \right]}{E \left[\tilde{S}(t | \overline{D}(t),Z, \bfL )\exp\left\{\int_{0}^{t}D(s)dB(s) \right\}  | Z, \bfL  \right]}. 
\end{eqnarray*}
Consequently, 
\begin{eqnarray*}
	&&\lambda^0(t |  \bfL) dt E \left[\tilde{S}(t | \overline{D}(t),Z, \bfL ) \exp\left\{\int_{0}^{t}D(s)dB(s) \right\}  |  Z, \bfL \right] \\
	&&=   E\left[ \tilde{S}(t | \overline{D}(t),Z, \bfL ) \exp\left\{\int_{0}^{t}D(s)dB(s) \right\}  \left\{ d\tilde{N}(t) -  D(t)dB(t) \right\} |  Z, \bfL   \right],
\end{eqnarray*}
with $\tilde{N}(t) = I(\tilde{T}\leq t)$.
Therefore, it holds that
\begin{eqnarray*}
	E \left[\tilde{Y}(t)  \exp\left\{\int_{0}^{t}D(s)dB(s) \right\}  \left\{ d\tilde{N}(t) - D(t)dB(t)
	 - \lambda^0(t |  \bfL) dt   \right\} |  Z, \bfL   \right] = 0. 
\end{eqnarray*} 
Consequently, 
\begin{equation}
	E \left[ \left\{Z-E(Z | \tilde{T}(0) \geq t, \bfL   ) \right\} \exp\left\{\int_{0}^{t}D(s)dB(s) \right\}  \left\{d\tilde{N}(t) - \tilde{Y}(t)D(t)dB(t) - \tilde{Y}(t)\lambda^0(t |  \bfL) dt \right\} |  \bfL \right] = 0.
	\label{eq: appendix conditional est eq}
\end{equation}
In the probability $E(Z | \tilde{T}(0) \geq t, \bfL)$, we condition on the counterfactual survival time under treatment, i.e. $\tilde{T}(0)$, to mimic the estimating equations of the Aalen least squares estimator (see Appendix \ref{Section: Appendix Score equations Aalen additive hazards model}) that also uses at-risk sets.

In the setting with right-censored data, it then follows  from (\ref{eq: appendix conditional est eq}) and the assumption $C\independent (\tilde{T},Z, \overline{D}(t)) |\bfL $ that 
\begin{eqnarray*}
&& 	E \left[ \left\{Z-E(Z | \tilde{T}(0) \geq t, \bfL   ) \right\} \exp\left\{\int_{0}^{t}D(s)dB(s) \right\}  \left\{d{N}(t) - {Y}(t)D(t)dB(t) - {Y}(t)\lambda^0(t |  \bfL) dt \right\} \right] \label{eq: appendix est eq observed data} \\
&&= 	E \left[ \left\{Z-E(Z | \tilde{T}(0) \geq t, \bfL   )  \right\} I(C\geq t)    \exp\left\{\int_{0}^{t}D(s)dB(s) \right\}  \right. \nn \\
&& \times\left. \left\{d\tilde{N}(t) - \tilde{Y}(t)D(t)dB(t) - \tilde{Y}(t)\lambda^0(t |  \bfL) dt \right\} \right] \nn \\
&&= 	E \left[   E(I(C\geq t )| \bfL)   \right. \nn \\
&&\left.  \times  E\left( \left\{Z-E(Z | \tilde{T}(0) \geq t, \bfL   ) \right\}  \exp\left\{\int_{0}^{t}D(s)dB(s) \right\}  \left\{d\tilde{N}(t) - \tilde{Y}(t)D(t)dB(t) - \tilde{Y}(t)\lambda^0(t |  \bfL) dt \right\}   | \bfL\right) \right] \nn \\
&& = 0 \nn, 
\end{eqnarray*}
which completes the proof of Theorem \ref{theorem: Estimating equation}.

\qedwhite

\subsection{One-step estimator for $\beta$ in model (\ref{eq: SNCSTM general})}
\label{Section: Appendix One-step estimator for constant treatment effect }

Under the SNCSTM assuming a constant treatment effect where patients can only cross over from the control to the experimental arm, i.e. 
\begin{align*}
	\frac{P[\tilde{T}(1_m, 0)>t  | \overline{D}(t_m)= \overline{1}, Z=1, \bfL, \tilde{T}\geq t_m]}{ P[\tilde{T}(1_{m-1}, 0)>t  | \overline{D}(t_m)= \overline{1}, Z=1, \bfL, \tilde{T}\geq t_m]} 
= \exp\left\{-  \beta  (t\wedge t_{m+1} - t_m) \right\},
\end{align*}
it holds that 
	\begin{equation}
		E \left[ \left\{Z-E(Z | \tilde{T}(0) \geq t, \bfL   ) \right\} \exp\left\{\beta \int_{0}^{t}D(s)ds \right\}  \left\{dN(t) -Y(t)D(t)\beta dt - Y(t)\lambda(t | Z=0, \bfL)dt\right\} \right] = 0, 
			\label{eq: appendix estimating equation constant}
	\end{equation}
	with 
	\begin{equation*}
		E(Z | \tilde{T}(0) \geq t, \bfL)	= \frac{E\left(ZY(t) \exp\left\{\beta\int_{0}^{t}D(s)ds \right\}  |\bfL \right)  }{E\left(Y(t) \exp\left\{\beta\int_{0}^{t}D(s)ds \right\} |\bfL  \right)}, 
	\end{equation*}
assuming $Z \independent \tilde{T}(0) | \bfL$  and $C\independent (\tilde{T},Z, \overline{D}(t)) |\bfL $. This follows from  Theorem	\ref{theorem: Estimating equation}, with $B(t) \equiv \beta t$ and acknowledging that the counterfactual hazard $\lambda^0(t |  \bfL)$ equals the observed hazard in the experimental arm $\lambda(t | Z=0, \bfL)$ as patients can not cross over from the experimental to the control arm. 
	
	Consequently, if $\tau$ is the time point indicating the end of the study, $\beta$ can be estimated by solving estimating equations 
	\begin{eqnarray}
		\frac{1}{n} \sum_{i=1}^{n} 	\bfU(Z_i,T_i,\overline{D_i}(\tau),\bfL_i;\beta) = 0, 
		\label{eq: app estimating equations method 1}
	\end{eqnarray}
	with 
	\begin{eqnarray}
		&&\bfU(Z,T,\overline{D}(\tau),\bfL;\beta) \nn  \\ 
		&&=   \int_{0}^{\tau} \left\{Z-	\hat{E}(Z | \tilde{T}(0) \geq t, \bfL) \right\} \exp\left\{\beta \int_{0}^{t}D(s)ds \right\} \nn \\
		&&\times \left\{ dN(t) -  Y(t)D(t)\beta dt - Y(t)\lambda(t | Z=0, \bfL)dt  \right\},  	\label{eq: appendix estimating equation}
	\end{eqnarray}
	where $	\hat{E}(Z | \tilde{T}(0) \geq t, \bfL)$ is an estimate of ${E}(Z | \tilde{T}(0) \geq t, \bfL)$ obtained by fitting a logistic regression model, with $Z$ as outcome,  baseline covariates $\bfL$ as predictors and weights $Y(t)\exp\left\{\beta \int_{0}^{t}D(s)ds \right\}$.  
	
	\subsubsection{Derivative of score equation}
	\label{Section: appendix Derivative of score equation}
	
	In this subsection, we calculate the derivative  of score equation  (\ref{eq: appendix estimating equation}) w.r.t. $\beta$.

	Let $\bfL = (1, L_1, \dots, L_{p-1})'$ denote a vector with constant 1,  baseline covariates and possible interactions between them.  
		If ${E}(Z | \tilde{T}(0) \geq t, \bfL)$  is estimated by fitting a logistic regression model, with $Z$ as outcome,   $\bfL$ as predictors and weights $Y(t)\exp\left\{\beta \int_{0}^{t}D(s)ds \right\}$, it holds that  
	\begin{equation}
		{E}(Z | \tilde{T}(0) \geq t, \bfL) = \expit\left(\bfgamma_t(\beta)'\bfL \right), 
		\label{eq: logistic model}
	\end{equation}
	with $\bfgamma_t(\beta)$ a parametric function of dimension $p \times 1$. This function is a function of $\beta$ as the   weights used in the estimating procedure depend on this parameter. 
	 From (\ref{eq: logistic model})  it follows that the derivative of the score equation (\ref{eq: appendix estimating equation}) w.r.t. $\beta$ equals
	\begin{eqnarray}
	&&\dot{\bfU}(Z,T,\overline{D}(\tau),\bfL;\beta)\nn \\
	&&=  \int_{0}^{\tau} \frac{\partial}{\partial \beta} (-{E}(Z | \tilde{T}(0) \geq t, \bfL))\exp\left\{\beta \int_{0}^{t}D(s)ds \right\} \left\{ dN(t) -  Y(t)D(t)\beta dt - Y(t)\lambda(t | Z=0, \bfL)dt  \right\} \nn \\
	&& +  \{Z- {E}(Z | \tilde{T}(0)\geq t, \bfL)\} \exp\left\{\beta \int_{0}^{t}D(s)ds \right\}  \left(\int_{0}^{t}D(s)ds  \right) \nn \\
	&& \times \left\{ dN(t) -  Y(t)D(t)\beta dt - Y(t)\lambda(t | Z=0, \bfL)dt  \right\} \nn \\ 
	&&-  \{Z- {E}(Z | \tilde{T}(0)\geq t, \bfL)\}  \exp\left\{\beta \int_{0}^{t}D(s)ds \right\}  Y(t)D(t) dt  \nn  \\
	&&=  \int_{0}^{\tau}- \frac{\expit\left(\bfgamma_t(\beta)'\bfL \right)}{1+\exp\left(\bfgamma_t(\beta)'\bfL \right)} \frac{\partial \bfgamma_t(\beta)}{\partial \beta}' \bfL  \exp\left\{\beta \int_{0}^{t}D(s)ds \right\}  \left\{ dN(t) -  Y(t)D(t)\beta dt - Y(t)\lambda(t | Z=0, \bfL)dt  \right\} \nn  \\
	&&+  \{Z-\expit\left(\bfgamma_t(\beta)'\bfL \right) \}   \exp\left\{\beta \int_{0}^{t}D(s)ds \right\}  \nn \\
	&& \times \left(\left( \int_{0}^{t}D(s)ds \right) \left\{ dN(t) -  Y(t)D(t)\beta dt  - Y(t)\lambda(t | Z=0, \bfL)dt \right\}  -  Y(t)D(t) dt\right). 
	\label{eq: appendix derivative estimating equation}
	\end{eqnarray}
The derivative $\partial \bfgamma_t(\beta)/\partial \beta$ can be obtained as follows. From the estimating equations used to estimate a logistic regression model with $Z$ as outcome,   $\bfL$ as predictors and weights $Y(t)\exp\left\{\beta \int_{0}^{t}D(s)ds \right\}$, 
it follows that 
		\begin{equation*}
		E \left[\bfL Y(t) \exp\left\{\beta \int_{0}^{t}D(s)ds \right\} \left\{Z- \expit\left(\bfgamma_t(\beta)'\bfL \right)\right\} \right] = 0,
	\end{equation*}
for all $\beta$. Because this is a constant function, then by 
 differentiating with respect to $\beta$, it follows that 
		\begin{eqnarray*}
		&&E \left[\bfL Y(t) \exp\left\{\beta \int_{0}^{t}D(s)ds \right\} \left( \int_{0}^{t}D(s)ds \right)\left\{Z- \expit\left(\bfgamma_t(\beta)'\bfL \right)\right\} \right] \\
	   &&- 	E \left[ \bfL Y(t) \exp\left\{\beta \int_{0}^{t}D(s)ds \right\}  \frac{\expit\left(\bfgamma_t(\beta)'\bfL \right)}{1+\exp\left(\bfgamma_t(\beta)'\bfL \right)} \frac{\partial \bfgamma_t(\beta)}{\partial \beta}' \bfL \right]  = 0.
	\end{eqnarray*}
Consequently, 
	\begin{eqnarray*}
	\frac{\partial \bfgamma_t(\beta)}{\partial \beta} =  &&E \left[ \bfL Y(t) \exp\left\{\beta \int_{0}^{t}D(s)ds \right\}  \frac{\expit\left(\bfgamma_t(\beta)'\bfL \right)}{1+\exp\left(\bfgamma_t(\beta)'\bfL \right)} \bfL' \right]^{-1} \\
	&& \times  E \left[\bfL Y(t) \exp\left\{\beta \int_{0}^{t}D(s)ds \right\} \left( \int_{0}^{t}D(s)ds \right)\left\{Z- \expit\left(\bfgamma_t(\beta)'\bfL \right)\right\} \right]. 
\end{eqnarray*}

\subsubsection{One-step estimator}

A one-step estimator \cite{le1956asymptotic} for $\beta$ can be obtained as follows.  Suppose $\hat{\beta}_0$ is an initial estimate for $\beta$. This estimator can be updated using a single Newton step: 
	\begin{equation*}
		\hat{\beta} = \hat{\beta}_0 - \sum_{i=1}^n {U}(\hat{\beta}_0, T_i, Z_i, \bfL_i, \overline{D}_i(t_M))/  \sum_{i=1}^n \dot{U}(\hat{\beta}_0, T_i, Z_i, \bfL_i, \overline{D}_i(t_M)), 
	\end{equation*}
 with ${U}(\hat{\beta}_0, T, Z, \bfL, \overline{D}(t_M))$ the score function, evaluated in $\hat{\beta}_0$, and $\dot{U}(\hat{\beta}_0, T, Z, \bfL, \overline{D}(t_M))$ the derivative of the score function w.r.t. $\beta$, evaluated in $\hat{\beta}_0$.

\subsection{Proof of double robustness}
\label{Section: Appendix Proof of double robustness}

\begin{theorem}
	\label{Theorem doubly robustness}
	The  estimator for $\beta$ discussed in the main paper and  Appendix \ref{Section: Appendix One-step estimator for constant treatment effect },  is  doubly robust, meaning that it is unbiased even if 
	the model for the randomized treatment, i.e. $E(Z | \tilde{T}(0) \geq t, \bfL)$, or the model for the hazards, i.e. $\lambda(t|  Z=0, \bfL)$, but not both,
	is misspecified.
\end{theorem}

\begin{lem}
	
Under SNCSTM (\ref{eq: appendix SNCSTM general}), 
	\begin{equation*}
	E \left[ \left\{Z-E(Z | \tilde{T}(0) \geq t, \bfL   ) \right\} \exp\left\{\beta \int_{0}^{t}D(s)ds \right\}  \left\{dN(t) -\beta Y(t)D(t)dt \right\} \right] = 0, 
\end{equation*}
with 
\begin{equation*}
	E(Z | \tilde{T}(0) \geq t, \bfL) = \frac{E\left[Z Y(t)\exp\left\{\beta \int_{0}^{t}D(s)ds \right\} |\bfL  \right]}{E\left[Y(t)\exp\left\{ \beta \int_{0}^{t}D(s)ds \right\} |\bfL  \right]}. 
\end{equation*}
assuming $Z \independent \tilde{T}(0) | \bfL$  and $C\independent (\tilde{T},Z, \overline{D}(t)) |\bfL $.
\label{lemma: estimating eq without hazard}
\end{lem}

\noindent \textbf{Proof}

\noindent Let $\omega(t | \bfL)$ be a function conditional on the baseline covariates $\bfL$. From equation (\ref{eq: survival T0 counterfactual}) it follows that 
\begin{eqnarray*}
	&&	E \left[ \left\{Z-E(Z | \tilde{T}(0) \geq t, \bfL   ) \right\}  \exp\left\{\beta \int_{0}^{t}D(s)ds \right\}  \tilde{Y}(t)\omega(t | \bfL) dt   |  \bfL  \right] \\
	&&= E \left[  \left\{Z-E(Z |  \tilde{T}(0) \geq t, \bfL   ) \right\} E\left( \tilde{Y}(t)   \exp\left\{\beta \int_{0}^{t}D(s)ds \right\} |  Z, \bfL \right)   \omega(t | \bfL)dt     |  \bfL    \right] \\
	&&= E \left[  \left\{Z-E(Z |  \tilde{T}(0) \geq t, \bfL   ) \right\} E\left( I(\tilde{T}(0) \geq t)    |  Z, \bfL \right)   \omega(t | \bfL) dt  |  \bfL    \right] \\
	&&= E \left[  \left\{Z-E(Z |  \tilde{T}(0) \geq t, \bfL   ) \right\}  |  \bfL    \right] E\left( I(\tilde{T}(0) \geq t)    |  \bfL \right)   \omega(t | \bfL) dt  \\
	&&= E \left[  \left\{Z-E(Z |  \tilde{T}(0) \geq t, \bfL   ) \right\} I(\tilde{T}(0) \geq t)   |  \bfL    \right]  \omega(t | \bfL) dt  \\
	&&= E \left[  \left\{Z-E(Z |  \tilde{T}(0) \geq t, \bfL   ) \right\}    |   \tilde{T}(0) \geq t, \bfL    \right]  P(\tilde{T}(0) \geq t |   \bfL )  \omega(t | \bfL) dt   =0,
\end{eqnarray*}
assuming $Z\independent  \tilde{T}(0)  |   \bfL $.  Consequently, from (\ref{eq: appendix conditional est eq}) it follows that
\begin{equation*}
	E \left[ \left\{Z-E(Z | \tilde{T}(0) \geq t, \bfL   ) \right\} \exp\left\{\beta \int_{0}^{t}D(s)ds \right\}  \left\{d\tilde{N}(t) - \beta \tilde{Y}(t)D(t) dt  \right\} |  \bfL \right] = 0.
\end{equation*}
In the setting with right-censored data, it then follows  from the assumption $C\independent (\tilde{T},Z, \overline{D}(t)) |\bfL $ that 
\begin{eqnarray*}
	&& 	E \left[ \left\{Z-E(Z | \tilde{T}(0) \geq t, \bfL   ) \right\}  \exp\left\{\beta \int_{0}^{t}D(s)ds \right\}  \left\{d{N}(t) -  \beta{Y}(t)D(t)dt \right\} \right] \\
	&&= 	E \left[ \left\{Z-E(Z | \tilde{T}(0) \geq t, \bfL   )  \right\} I(C\geq t)    \exp\left\{\beta \int_{0}^{t}D(s)ds \right\}  \left\{d\tilde{N}(t) -\beta \tilde{Y}(t)D(t)dt \right\} \right] \nn \\
	&&= 	E \left[   E(I(C\geq t )| \bfL)   \right. \nn \\
	&&\left.  \times  E\left( \left\{Z-E(Z | \tilde{T}(0) \geq t, \bfL   ) \right\}  \exp\left\{\beta \int_{0}^{t}D(s)ds \right\}  \left\{d\tilde{N}(t) - \beta \tilde{Y}(t)D(t)dt \right\}   | \bfL\right) \right] \nn \\
	&& = 0.  \nn
\end{eqnarray*}

\qedwhite

\noindent \textbf{Proof of Theorem \ref{Theorem doubly robustness}}

\noindent First, let $\lambda^*(t|  Z=0, \bfL)$ be an estimator for $\lambda(t|  Z=0, \bfL)$, which is possibly misspecified. We will prove that equation (\ref{eq: appendix estimating equation constant}) is still on average 0.  Equation (\ref{eq: appendix estimating equation constant})  can be rewritten as 
\begin{eqnarray*}
	&& E \left[ \left\{Z-E(Z | \tilde{T}(0) \geq t, \bfL   ) \right\} \exp\left\{\beta \int_{0}^{t}D(s)ds \right\}  \left\{dN(t) -Y(t)D(t)\beta dt - Y(t)\lambda^*(t | Z=0, \bfL)dt\right\} \right] \\ 
	&&=  E \left[ \left\{Z-E(Z | \tilde{T}(0) \geq t, \bfL   ) \right\} \exp\left\{\beta \int_{0}^{t}D(s)ds \right\}  \left\{dN(t) -Y(t)D(t)\beta dt \right\} \right] \\ 
	&&- E \left[ \left\{Z-E(Z | \tilde{T}(0) \geq t, \bfL   ) \right\} \exp\left\{\beta \int_{0}^{t}D(s)ds \right\}  Y(t)\lambda^*(t | Z=0, \bfL)dt \right] \\
	&&= - E \left[ \left\{Z-E(Z | \tilde{T}(0) \geq t, \bfL   ) \right\} E\left(\exp\left\{\beta \int_{0}^{t}D(s)ds \right\}  Y(t)  | Z, \bfL\right)\lambda^*(t | Z=0, \bfL)dt \right], 
\end{eqnarray*}
where the second equality follows from Lemma \ref{lemma: estimating eq without hazard}.  Using equality (\ref{eq: survival T0 conditional}) and assuming  $ Z \independent \tilde{T}(0) |  \bfL $, this can be further rewritten as 
\begin{eqnarray*}
&&- E \left[ \left\{Z-E(Z | \tilde{T}(0) \geq t, \bfL   ) \right\} E\left( I(\tilde{T}(0) \geq t)  | Z, \bfL\right) P(C \geq t | \bfL) \lambda^*(t | Z=0, \bfL)dt \right] \\
&&= - E \left[ \left\{Z-E(Z | \bfL   ) \right\} E\left( I(\tilde{T}(0) \geq t)  |  \bfL\right) P(C \geq t | \bfL) \lambda^*(t | Z=0, \bfL)dt \right] \\
&&= - E \left[ E\left( \left\{Z-E(Z | \bfL   ) \right\} |  \bfL\right)E\left( I(\tilde{T}(0) \geq t)  |  \bfL\right) P(C \geq t | \bfL) \lambda^*(t | Z=0, \bfL)dt \right] \\
&& =0. 
\end{eqnarray*}

Next, let $E^*(Z | \tilde{T}(0) \geq t, \bfL   )$ be an estimator for $E(Z | \tilde{T}(0) \geq t, \bfL   )$, which is possibly misspecified. From equation (\ref{eq: survival T0 counterfactual}) and the assumption $Z \independent \tilde{T}(0) | \bfL$, it follows that 
\begin{equation*}
	P(\tilde{T}(0) \geq t |\bfL) =  E\left(\tilde{Y}(t)\exp\left\{\beta \int_{0}^{t}D(s)ds \right\} |Z, \bfL \right). 
\end{equation*}
By taking logs of each side of this equation and differentiating with respect to $t$, it follows that 
\begin{eqnarray*}
	\lambda^0(t |  \bfL) 
	&&= - \frac{d}{dt} \log E \left[\tilde{Y}(t) \exp\left\{\beta \int_{0}^{t}D(s)ds \right\}  |  Z, \bfL \right].
\end{eqnarray*}
 It then follows that
\begin{eqnarray*}
	&&\lambda^0(t |  \bfL) \\
	&&= - \frac{d}{dt} \log E \left[\tilde{S}(t | \overline{D}(t),Z, \bfL )\exp\left\{\beta \int_{0}^{t}D(s)ds \right\} |  Z, \bfL \right]\\ 
	&&= - \frac{ E\left[-\tilde{f}(t |\overline{D}(t),Z, \bfL ) \exp\left\{\beta \int_{0}^{t}D(s)ds \right\} +\tilde{S}(t | \overline{D}(t),Z, \bfL ) \frac{d}{dt} \left(  \exp\left\{\beta \int_{0}^{t}D(s)ds \right\} \right) |  Z, \bfL  \right]}{E \left[\tilde{S}(t | \overline{D}(t),Z, \bfL)\exp\left\{\beta \int_{0}^{t}D(s)ds \right\} |  Z, \bfL \right]}.
\end{eqnarray*}
It follows that 
\begin{eqnarray*}
	\lambda^0(t |  \bfL) 
	&&=  \frac{ E\left[ \tilde{S}(t |\overline{D}(t),Z, \bfL)\exp\left\{\beta \int_{0}^{t}D(s)ds \right\}  \left\{\frac{\tilde{f}(t | \overline{D}(t),Z, \bfL )}{\tilde{S}(t | \overline{D}(t),Z, \bfL )} - D(t)\beta \right\} |  Z, \bfL   \right]}{E \left[\tilde{S}(t | \overline{D}(t),Z, \bfL )\exp\left\{\beta \int_{0}^{t}D(s)ds \right\}  | Z, \bfL  \right]}. 
\end{eqnarray*}
Consequently, 
\begin{eqnarray*}
	&&\lambda^0(t |  \bfL) dt E \left[\tilde{S}(t | \overline{D}(t),Z, \bfL ) \exp\left\{\beta \int_{0}^{t}D(s)ds \right\}  |  Z, \bfL \right] \\
	&&=   E\left[ \tilde{S}(t | \overline{D}(t),Z, \bfL )\exp\left\{\beta \int_{0}^{t}D(s)ds \right\}  \left\{ d\tilde{N}(t) -  \beta D(t)dt \right\} |  Z, \bfL   \right].
\end{eqnarray*}
Therefore, it holds that
\begin{eqnarray*}
	E \left[\tilde{Y}(t) \exp\left\{\beta \int_{0}^{t}D(s)ds \right\}  \left\{ d\tilde{N}(t) - \beta D(t)dt
	-\lambda(t|  Z=0, \bfL) dt   \right\} |  Z, \bfL   \right] = 0,  
\end{eqnarray*} 
with $\lambda(t|  Z=0, \bfL) = \lambda^0(t |  \bfL) $ as patients in the experimental arm stay on experimental treatment.  
In the setting with right-censored data, it then follows  from the assumption $C\independent (\tilde{T},Z, \overline{D}(t)) |\bfL $ that 
\begin{eqnarray*}
&&E \left[{Y}(t) \exp\left\{\beta \int_{0}^{t}D(s)ds \right\}  \left\{ d{N}(t) - \beta D(t)dt
-\lambda(t|  Z=0, \bfL) dt   \right\} |  Z, \bfL   \right] \\
&& =   E\left[I(C\geq t ) |  Z, \bfL \right]	E \left[\tilde{Y}(t) \exp\left\{\beta \int_{0}^{t}D(s)ds \right\}  \left\{ d\tilde{N}(t) - \beta D(t)dt
-\lambda(t|  Z=0, \bfL) dt   \right\} |  Z, \bfL   \right] \\
&& = 0. 
\end{eqnarray*} 
Consequently, 
\begin{eqnarray*}
	&& E \left[ \left\{Z-E^*(Z | \tilde{T}(0) \geq t, \bfL   ) \right\} \exp\left\{\beta \int_{0}^{t}D(s)ds \right\}  \left\{dN(t) -Y(t)D(t)\beta dt - Y(t)\lambda(t | Z=0, \bfL)dt\right\} \right] \\ 
	&& = E \left[ \left\{Z-E^*(Z | \tilde{T}(0) \geq t, \bfL   ) 	\right\}  \right.\\
&& \left. \times E\left(\exp\left\{\beta \int_{0}^{t}D(s)ds \right\}  \left\{dN(t) -Y(t)D(t)\beta dt - Y(t)\lambda(t | Z=0, \bfL)dt\right\}  |  Z, \bfL  \right) \right] \\
	&&=0.
\end{eqnarray*}
\qedwhite

\newpage
\section{Counterfactual survival curve under control}
\label{Section: Appendix counterfactual survival curve under control}

If patients can only cross over from the control arm to the experimental arm and not vice versa, the survival probability if all patients would take control treatment and not cross over can be estimated as follows. From the hypothetical risk ratio (\ref{eq: appendix hypothetical survival risk ratio}) and assumption $Z \independent \tilde{T}(0) | \bfL$ it follows that 
\begin{eqnarray}
  P(\tilde{T}(1)>t |\bfL ) 
	&&= \exp\{-B(t)\} P(\tilde{T}(0)>t |\bfL ) \nn \\
	&&= \exp\{-B(t)\} P(\tilde{T}(0)>t | Z = 0, \bfL) \nn \\
	&&= \exp\{-B(t)\} P(\tilde{T}>t | Z = 0, \bfL), 
	\label{eq: appendix counterfactual control survival}
\end{eqnarray}
where the third equality follows since all treated patients take experimental treatment for the entire duration of the trial. 
Marginal survival probabilities can be obtained by averaging out the baseline covariates $\bfL$ in equation (\ref{eq: appendix counterfactual control survival}): 
\begin{eqnarray*}
	P(\tilde{T}(1)>t ) 
	&&= \exp\{-B(t)\} E\left[P(\tilde{T}>t | Z = 0, \bfL )\right] . \label{eq: appendix marginal counterfactual control survival}
\end{eqnarray*}
Therefore, the marginal probabilities can be obtained by averaging over all patients in the dataset:
\begin{eqnarray*}
	\hat{P}(\tilde{T}(1)>t ) 
	&&= \exp\{-\hat{B}(t)\} \frac{1}{n} \sum_{i=1}^{n}  \hat{P}(\tilde{T}>t | Z_i = 0, \bfL_i ). 
\end{eqnarray*}
The survival probabilities $\hat{P}(\tilde{T}>t | Z_i = 0, \bfL_i )$ can be obtained by fitting a Cox proportional hazards model or an Aalen additive hazards model to the observed data. 

\newpage
\section{Data analysis}
\label{Section: Appendix Data analysis }

\subsection{Estimands framework}
\label{Section: Appendix Data analysis guideline}

The treatment-policy and hypothetical estimand in the HELIOS trial, described in section \ref{Section: Data analysis}, can be defined according to the estimands framework described in the addendum of the ICH E9(R1) guideline \cite{ICH2019}: 

\begin{itemize}
\item \textbf{Treatment}: 420mg daily ibrutinib or placebo, plus 6 cycles of bendamustine plus rituximab, as defined by the study protocol. 

\item \textbf{Population}: the entire study population, as defined by the inclusion-exclusion criteria of the study. 

\item \textbf{Variable}: time to death. 

\item \textbf{Intercurrent events}: crossover from the control  to the experimental treatment: 

\begin{itemize}
	\item \textbf{Treatment-policy estimand}: the value of the outcome is of interest regardless of crossover.
	
	\item \textbf{Hypothetical estimand}: the hypothetical scenario is envisaged where patients on the control arm would not start experimental treatment. 
\end{itemize}

\item \textbf{Population-level summary}:  the  risk ratio comparing survival times in the experimental arm versus the control arm. 
\begin{itemize}
   \item \textbf{Treatment-policy estimand}: the  risk ratio at the end of the trial, contrasting the survival probabilities if all patients would be assigned to the experimental treatment arm with the survival probabilities if all patients would be assigned to the control arm. 
    \item \textbf{Hypothetical estimand}: the  risk ratio at the end of the trial, contrasting the survival probabilities if all patients would take experimental treatment for the entire duration of the trial with the survival probabilities if all patients would take control treatment for the entire duration of the trial.
\end{itemize}
Here, we only considered the intercurrent event ‘crossover from the control  to the experimental treatment’ and handled it using the treatment-policy or hypothetical estimand. However, in clinical trials different intercurrent events might occur, such as starting prohibited medication. According to the addendum of the ICH E9(R1) guideline, it is not necessary to use the same estimand to address all intercurrent events. In particular, different estimands can be used to reflect the clinical question of interest in respect to different intercurrent events.

\end{itemize}

%

\subsection{Aalen additive hazards model}
\label{Section: Appendix Score equations Aalen additive hazards model}

We consider a semiparametric additive hazards model \cite{mckeague1994partly} with time-varying intercept $ \delta(t)$ and constant treatment effect $\beta_A$: 
\begin{equation*}
\lambda(t | Z) = Y(t)  \left\{   \delta(t) + \beta_A Z \right\}.  
\end{equation*}
To estimate $\beta$, we solve estimating equations 
\begin{eqnarray}
	\frac{1}{n} \sum_{i=1}^{n} 	\bfU(Z_i,T_i;\beta_A) = 0, 
	\label{eq: app estimating equations Aalen}
\end{eqnarray}
with 
\begin{eqnarray}
\bfU(Z,T;\beta_A) \nn  =   \int_{0}^{\tau} \left\{Z-	E(ZY(t))/E(Y(t)) \right\} \left\{ dN(t) -  Y(t)D(t)\beta_A dt  - Y(t)\lambda(t)dt \right\}.	\label{eq: appendix estimating equation Aalen}
\end{eqnarray}
To obtain a $p$-value for the test $\beta_A = 0$, a one-sample t-test is performed on the score functions $U(Z,T;\beta_A = 0) $ evaluated in $\beta_A = 0$. The variance of $\hat{\beta}_A$ is estimated using the  sandwich estimator. 
 In addition, a 95\% confidence interval for $\beta_A$ is found by searching for the values that lead to  a $p$-value of 5\% in the t-test on the score functions. 
 
 \subsection{Score equations IPCW}
 \label{Section: Appendix Score equations IPCW}
 
 Let $S$ denote the time to crossover for control treatments, defined as $\infty$ if the patient does not  cross over. In addition, let $\mathbf X(t)$ be a vector of longitudinal confounders for crossover. The stabilized inverse probability weights \cite{robins2000marginal} to correct for censoring for patient $i$ are then defined as: 
 \begin{eqnarray*}
 	W_{\text{stab},i}(t,\mathbf X(t),\bfL,S,T) = \frac{P(S>t | \bfL, T\geq t, S\geq t, Z=1)}{P(S>t |\mathbf X(t), \bfL, T\geq t, S\geq t, Z=1)}. 
 \end{eqnarray*}
 If only baseline covariates are included to predict crossover, the weights are defined as: 
  \begin{eqnarray*}
 	W_{\text{baseline},i}(t,\bfL,S,T) = \frac{1}{P(S>t |\bfL, T\geq t, S\geq t, Z=1)}. 
 \end{eqnarray*}
When IPCWeights are used to correct for crossover, $\beta_A$ is estimated by solving estimating equations 
\begin{eqnarray*}
	\frac{1}{n} \sum_{i=1}^{n} 	\bfU(Z_i,T_i;\beta_A) = 0, 
\end{eqnarray*}
with 
\begin{eqnarray*}
	\bfU(Z,T;\beta_A) \nn  =   \int_{0}^{\tau} W_i(t) \left\{Z-	E(ZY(t))/E(Y(t)) \right\} \left\{ dN(t) -  Y(t)D(t)\beta_A dt  - Y(t)\lambda(t)dt \right\},
\end{eqnarray*}
where $W_i(t) $ equals $W_{\text{stab},i}(t,\mathbf X(t),\bfL,S,T) $ or $W_{\text{baseline},i}(t,\bfL,S,T)$.

\subsection{Results}
\label{Section: Appendix Data analysis Results }

\begin{figure}[H]
	\includegraphics[width=\textwidth]{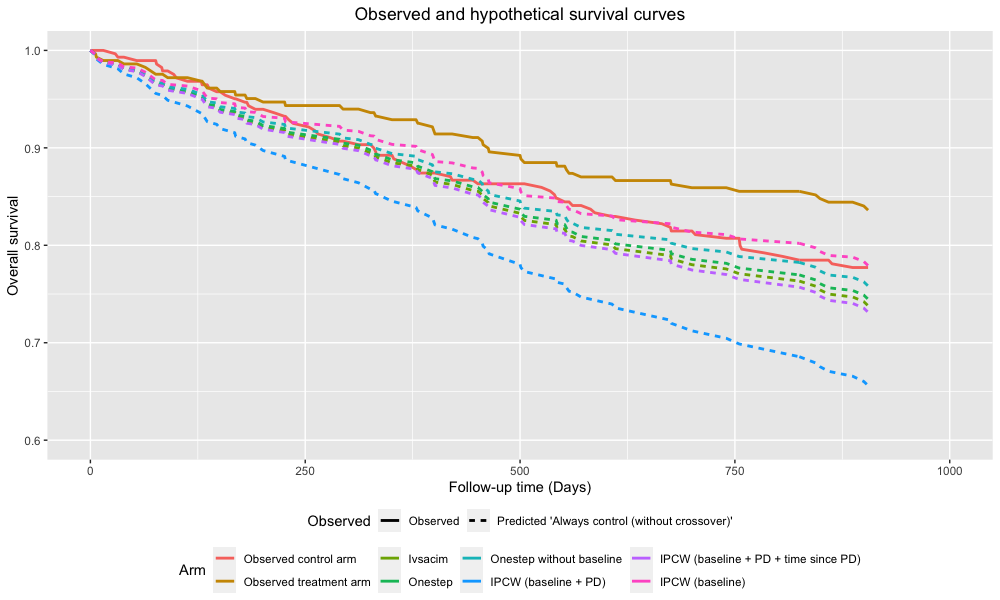} 
	\caption{Observed survival curves in the HELIOS trial, and predicted curves if all patients would take control for the entire duration of the trial, based on different methods.}
	\label{fig: Appendix Data analysis Results hypothetical survival curves}
\end{figure}

\subsubsection{Hypothetical estimand: One-step estimator}
\label{Section: Appendix Hypothetical estimand: Onestep estimator }

\begin{figure}[H]
	\includegraphics[width=\textwidth]{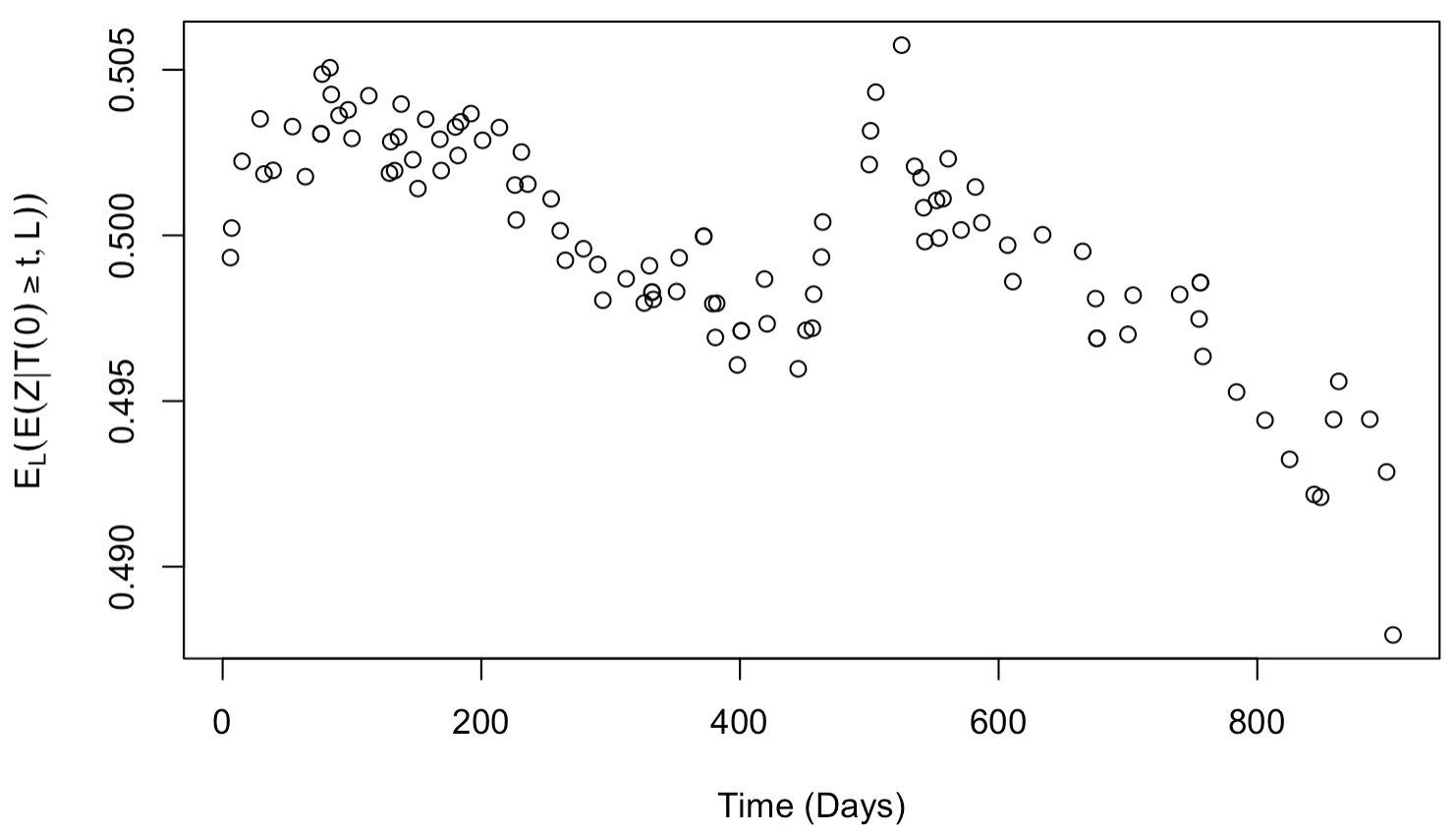} 
	\caption{Plot of the estimated $E_{\bfL}(\hat{E}(Z | \tilde{T}(0) \geq t, \bfL))$ for the obtained $\beta$ using the one-step estimator, in function of $t$.}
	\label{fig: Appendix Data analysis Results function beta}
\end{figure}

Under the assumption $Z \independent \tilde{T}(0) | \bfL$, ${E}(Z | \tilde{T}(0) \geq t, \bfL)$ does not depend on $t$. To verify this, $E_{\bfL}(\hat{E}(Z | \tilde{T}(0) \geq t, \bfL))$ was estimated for the obtained $\beta$ using the one-step estimator. Here, $E_{\bfL}$ indicates that the average was taken over the baseline covariates $\bfL$. In figure \ref{fig: Appendix Data analysis Results function beta}, it can be observed that $E_{\bfL}(\hat{E}(Z | \tilde{T}(0) \geq t, \bfL))$  indeed practically does not depend on $t$ in the HELIOS trial. 

%
%
%
%

\subsubsection{Hypothetical estimand: IPCW estimators}
\label{Section: Appendix Data analysis Results Hypothetical estimand: IPCW estimators}

\noindent \textbf{Only baseline covariates as predictors}

	\begin{table}[H]
	\centering
	\begin{tabular}{lccc}
		\toprule
		\textbf{Variable} & \textbf{Coefficient} &  \textbf{SE} & \textbf{$p$-value} \\ 
		\cmidrule(ll){1-4} 
		STRATA1 (1) & -0.1544   &    0.1663 &   0.353 \\
		STRATA2 (yes) & 0.3037    &   0.1968  &   0.123 \\
		SEX (man) & 0.1609 & 0.1750 & 0.358 \\
		AGE & -0.0034  &  0.0086 &      0.694 \\
		\bottomrule
	\end{tabular}
	\caption{Cox proportional hazards model for the crossover time in function of the baseline covariates. This model is used in the denominator of the inverse probability weights.  }
\end{table}

\begin{figure}[H]
	\includegraphics[width=0.9\textwidth]{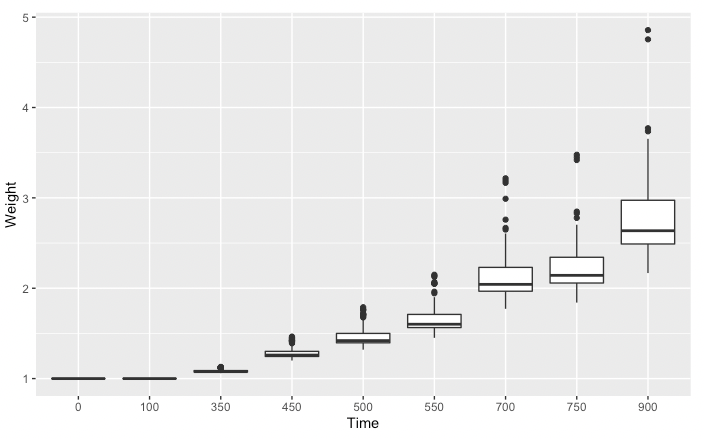} 
	\caption{Histogram of the  weights for the control patients used in the IPCW approach with only baseline predictors. \\(To make the plot  more clear, only weights per 50 days are shown) \\
		Min: 1.00, 1st quantile: 1.00, median: 1.073, mean: 1.33, 3rd quantile: 1.48, max: 4.97.}
	\label{fig: Appendix Data analysis Results Weights IPCW Baseline}
\end{figure}

\noindent \textbf{Baseline covariates and indicator for PD as predictors}

	\begin{table}[H]
	\centering
	\begin{tabular}{lccc}
		\toprule
		\textbf{Variable} & \textbf{Coefficient} &  \textbf{SE} & \textbf{$p$-value} \\ 
		\cmidrule(ll){1-4} 
		STRATA1 (1) & -0.1539   &    0.1663 &   0.355 \\
		STRATA2 (yes) & 0.3024    &   0.1968  &   0.124 \\
		SEX (man) & 0.1600 &   0.1750 &     0.361\\
		AGE &  -0.0034 &    0.0086 &      0.693 \\ 
		\bottomrule
	\end{tabular}
	\caption{Cox proportional hazards model for the crossover time in function of the baseline covariates. This model is used in the numerator of the stabilized inverse probability weights. }
\end{table}

\begin{table}[H]
	\centering
	\begin{tabular}{lccl}
		\toprule
		\textbf{Variable} & \textbf{Coefficient} &  \textbf{SE} & \textbf{$p$-value} \\ 
		\cmidrule(ll){1-4} 
		STRATA1 (1) & 0.3202 & 0.1776 &$ 0.0714^*$ \\
		STRATA2 (yes) &  0.0560 &   0.2093  &   0.7890  \\
		SEX (man) & 0.0057 & 0.1865 & 0.9754 \\
		AGE & 0.0129 & 0.0087 & 0.1357\\
		PD & 22.225  &  2.906 $\times 10^3$  &    0.9939   \\
		\bottomrule
	\end{tabular}
	\caption{Cox proportional hazards model for the crossover time in function of the baseline covariates and a time-varying indicator for disease progression (PD). This model is used in the denominator of the stabilized inverse probability weights. }
\end{table}

\begin{figure}[H]
	\includegraphics[width=\textwidth]{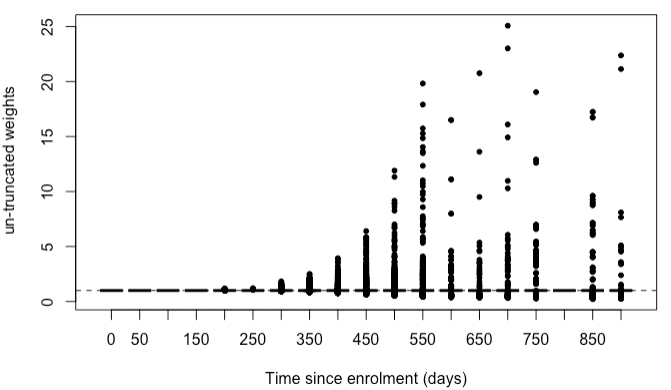} 
	\caption{Weight distribution of the stabilized weights for the control patients used in the IPCW approach with baseline and disease progression predictors.
		\\(To make the plot  more clear, only weights per 50 days are shown)  \\
	Min: 0.23, 1st quantile: 0.91, median: 1.00, mean: 1.02, 3rd quantile: 1.00, max: 25.07.}
	\label{fig: Appendix Data analysis Results Weights IPCW PD}
\end{figure}

\noindent \textbf{Baseline covariates, indicator for PD and time since PD as predictors}

	\begin{table}[H]
	\centering
	\begin{tabular}{lccc}
		\toprule
		\textbf{Variable} & \textbf{Coefficient} &  \textbf{SE} & \textbf{$p$-value} \\ 
		\cmidrule(ll){1-4} 
		STRATA1 (1) & -0.1539   &    0.1663 &   0.355 \\
		STRATA2 (yes) & 0.3024    &   0.1968  &   0.124 \\
		SEX (man) & 0.1600 &   0.1750 &     0.361\\
		AGE &  -0.0034 &    0.0086 &      0.693 \\ 
		\bottomrule
	\end{tabular}
	\caption{Cox proportional hazards model for the crossover time in function of the baseline covariates. This model is used in the numerator of the stabilized inverse probability weights. }
\end{table}

\begin{table}[H]
	\centering
	\begin{tabular}{lccl}
		\toprule
		\textbf{Variable} & \textbf{Coefficient} &  \textbf{SE} & \textbf{$p$-value} \\ 
		\cmidrule(ll){1-4} 
		STRATA1 (1) & 0.2025 &    0.1771 &   0.2528    \\
		STRATA2 (yes) &   0.1918   &  0.2092& 0.3594     \\
		SEX (man) & 0.0265 &  0.1828 &   0.8848   \\
		AGE &0.0071 &     0.0089&  0.4270 \\
		PD & 22.31  &  2.517$\times 10^3$&  0.9929\\
		Time since PD & -0.0036 &    0.0009&  $0.0001^{***}$ \\
		\bottomrule
	\end{tabular}
	\caption{Cox proportional hazards model for the crossover time in function of the baseline covariates and a time-varying indicator for disease progression (PD) and time since disease progression. This model is used in the denominator of the stabilized inverse probability weights. }
\end{table}

\begin{figure}[H]
	\includegraphics[width=\textwidth]{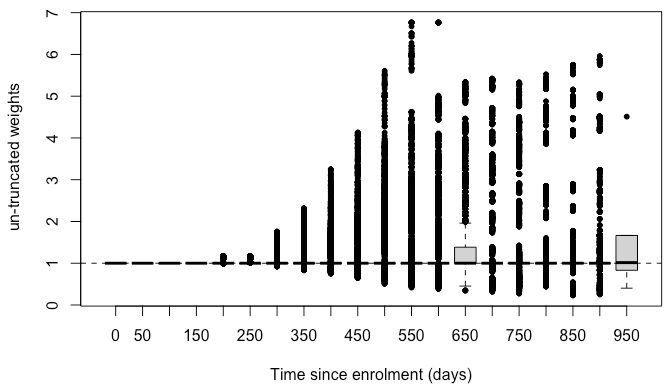} 
	\caption{Weight distribution of the stabilized inverse probability weights for the control patients used in the IPCW approach with baseline, indicator for disease progression and time since disease progression as predictors.  We did not find evidence for interactions or non-linearity. 
		\\(To make the plot  more clear, only weights per 50 days are shown) \\
		Min: 0.23, 1st quantile: 0.98, median: 1.00, mean: 1.41, 3rd quantile: 1.47, max: 6.76.}
	\label{fig: Appendix Data analysis Results Weights IPCW Time}
\end{figure}

%
%
%
%
%
%
%
%
%
%
%

\end{appendices}

\end{document}